\documentclass[preprint,12pt,review]{elsarticle}

\usepackage{amssymb}
\usepackage[font={small}]{caption}
\usepackage{times}
\usepackage{graphicx}
\usepackage{colortbl}
\usepackage{amsmath}
\usepackage{xcolor}
\usepackage{soul}
\usepackage{booktabs}
\usepackage[margin=1in]{geometry}

\journal{Additive Manufacturing}

\begin{document}

\begin{frontmatter}



\title{Limits of dispersoid size and number density in oxide dispersion strengthened alloys fabricated with powder bed fusion-laser beam}

\author[inst1]{Nathan A. Wassermann}

\affiliation[inst1]{organization={Department of Mechanical Engineering},
            addressline={5000 Forbes Ave}, 
            city={Pittsburgh},
            postcode={15213}, 
            state={Pittsburgh},
            country={USA}}

\author[inst2]{Yongchang Li}

\affiliation[inst2]{organization={Department of Nuclear Engineering},
            addressline={400 Bizzell St}, 
            city={College Station},
            postcode={77843}, 
            state={Texas},
            country={USA}}
            
\author[inst1]{Alexander J. Myers}

\author[inst3]{Christopher A. Kantzos}
\affiliation[inst3]{organization={High Temperature and Smart Alloy Branch, NASA Glenn Research Center},
            addressline={21000 Brookpark Rd}, 
            city={Cleveland},
            postcode={44135}, 
            state={Ohio},
            country={USA}}

\author[inst3]{Timothy M. Smith}
\author[inst1]{Jack L. Beuth}
\author[inst1]{Jonathan A. Malen}
\author[inst2]{Lin Shao}
\author[inst1]{Alan J.H. McGaughey}
\author[inst1]{Sneha P. Narra}

\begin{abstract}
Previous work on additively-manufactured oxide dispersion strengthened alloys focused on experimental approaches, resulting in larger dispersoid sizes and lower number densities than can be achieved with conventional powder metallurgy. To improve the as-fabricated microstructure, this work integrates experiments with a thermodynamic and kinetic modeling framework to probe the limits of the dispersoid sizes and number densities that can be achieved with powder bed fusion-laser beam. Bulk samples of a Ni-20Cr $+$ 1 wt.\% Y$_2$O$_3$ alloy are fabricated using a range of laser power and scanning velocity combinations. Scanning transmission electron microscopy characterization is performed to quantify the dispersoid size distributions across the processing space. The smallest mean dispersoid diameter (29 nm) is observed at 300 W and 1200 mm/s, with a number density of 1.0$\times$10$^{20}$ m$^{-3}$. The largest mean diameter (72 nm) is observed at 200 W and 200 mm/s, with a number density of 1.5$\times$10$^{19}$ m$^{-3}$. Scanning electron microscopy suggests that a considerable fraction of the oxide added to the feedstock is lost during processing, due to oxide agglomeration and the ejection of oxide-rich spatter from the melt pool. After accounting for these losses, the model predictions for the dispersoid diameter and number density align with the experimental trends. The results suggest that the mechanism that limits the final number density is collision coarsening of dispersoids in the melt pool. The modeling framework is leveraged to propose processing strategies to limit dispersoid size and increase number density.
\end{abstract}

\begin{graphicalabstract}
\includegraphics[width=1\linewidth]{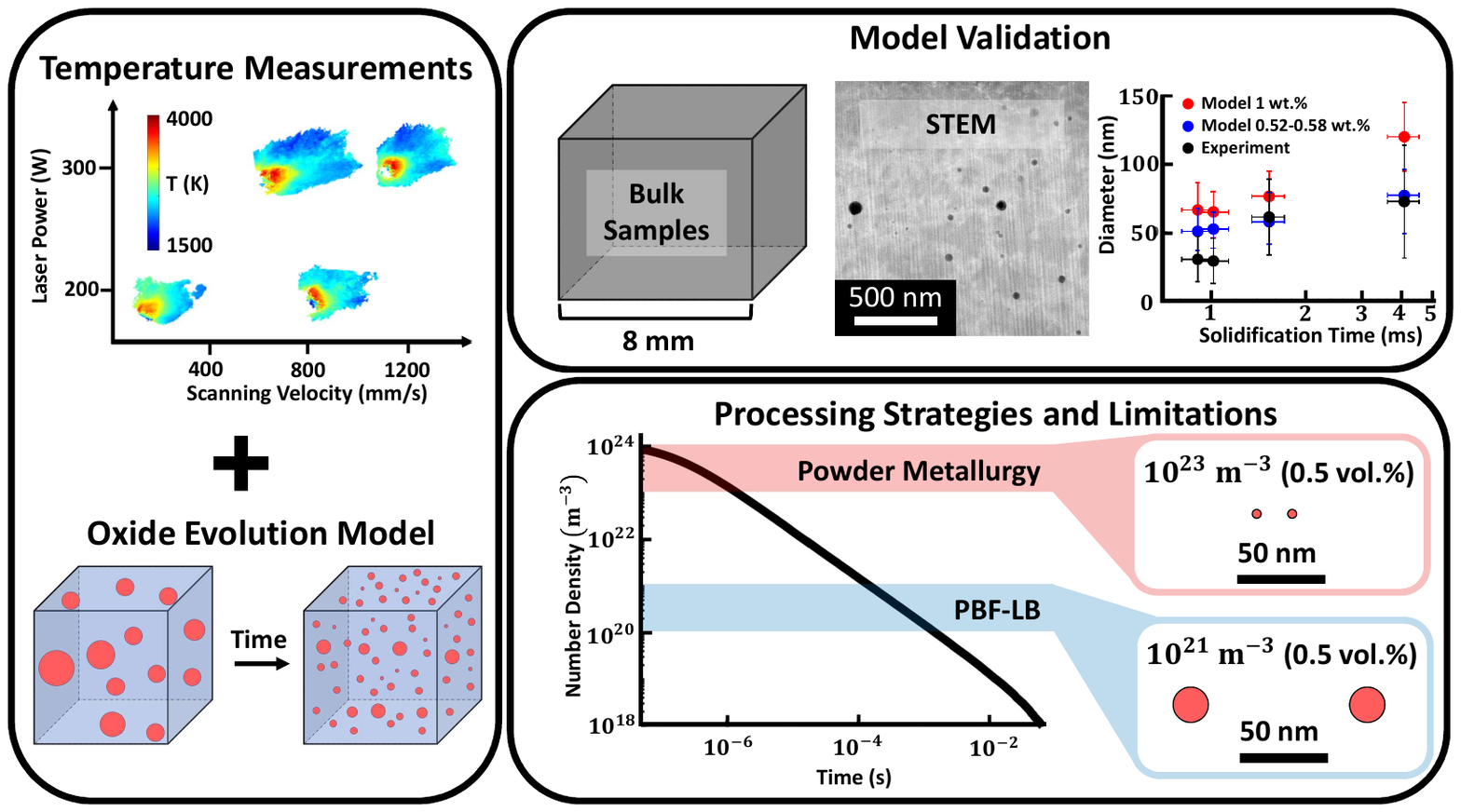}
\end{graphicalabstract}

\begin{highlights}
\item Bulk samples of a Ni-20Cr ODS alloy are fabricated with powder bed fusion-laser beam.
\item A model is applied to gain insight into the evolution of oxide dispersoids.
\item STEM characterization is performed across the processing space for model validation.
\item Shorter dispersoid residence times yield higher number densities and smaller sizes.
\item Number density is limited by collision coarsening of dispersoids in the melt pool.
\end{highlights}


\end{frontmatter}


\section{Introduction}
\label{intro}
Oxide dispersion strengthened (ODS) alloys offer exceptional creep strength at elevated temperatures (650 ℃ $-$ 1000 ℃), enabling high performance in extreme service environments such as aerospace engines \cite{nasa20, nasa23} and nuclear reactors \cite{zinkle, odette}. The performance of ODS alloys depends on the characteristics of the dispersoids (i.e., nanometric oxide particles), including their size and number density. For example, proposed nuclear fusion applications demand high number densities (10$^{23}$ m$^{-3}$ $-$ 10$^{24}$ m$^{-3}$) for helium management, with dispersoid diameters ranging from 2 nm to 5 nm \cite{odette}. On the other hand, optimum creep resistance is achieved with larger diameters ($\sim$20 nm) and lower number densities (10$^{21}$ m$^{-3}$) in ODS alloys designed for high-temperature environments \cite{rosler}.\par

Though ODS alloys promise a range of desirable properties, they have fallen out of the market due to the limited scalability of conventional powder metallurgy-based manufacturing techniques. In the first processing step, a highly thermodynamically-stable oxide (typically Y$_2$O$_3$) is dissolved into an alloy powder via mechanical alloying, a process that is defect-prone and energy-intensive. In the second step, high pressure is applied at sub-melting point temperatures to join the alloy powder and precipitate the oxide dispersoids. Since this process offers limited geometrical complexity, extensive thermomechanical post-processing is required to fabricate useful structural forms like tubes and claddings, resulting in a highly anisotropic microstructure \cite{odette}. Fabrication of ODS components with complex geometries presents additional challenges since ODS alloys cannot be joined using traditional arc welding processes, where oxide particles rise due to buoyancy and agglomerate on the top surface of the melt pool, eliminating their strengthening capability \cite{zinkle, odette}. To address these limitations, recent work has focused on additive manufacturing (AM) approaches \cite{nasa23, horn, Eo, zhong, kenel}, which enable the fabrication of near-net shape components through a layerwise process. The majority of these studies focused on powder bed fusion-laser beam (PBF-LB), which yields melt pool solidification times three orders of magnitude smaller than arc welding (depending on the material and processing conditions) \cite{hongFluid}, allowing less time for coarsening of the oxide dispersoids. These accelerated solidification times ($\sim$1 ms) have inspired optimism that PBF-LB could yield dispersoid number densities comparable to those achievable with powder metallurgy ($>$10$^{23}$ m$^{-3}$ \cite{klueh2002tensile}). \par 

Though PBF-LB presents advantages over the traditional manufacturing route, previous studies have highlighted several challenges, including low dispersoid number densities (typically 10$^{19}$ m$^{-3}$ $-$ 10$^{21}$ m$^{-3}$ \cite{nasa20, horn, Eo, haines, saptarshi}), oxide agglomeration \cite{zhong}, and slagging \cite{kenel, de2023HfO2}. An extensive review of the challenges associated with additive manufacturing of ODS alloys has been published by Wilms \textit{et al.}~\cite{wilms2023additive}. The previous studies relied almost exclusively on experimentation to test a variety of alloy compositions and processing conditions (e.g., laser power and scanning velocity). The feasibility of this approach is limited, however, since each experiment typically requires nanoscale characterization such as transmission electron microscopy (TEM) to accurately measure the size distribution and number density of nanometric dispersoids. Due to the need for time-consuming characterization, dispersoid characteristics are typically reported only for a single sample \cite{horn, kenel, saptarshi}. There is thus limited information about the trends in dispersoid size and number density across the laser power and scanning velocity processing space. Therefore, a modeling framework for the evolution of oxide nanoparticles during PBF-LB is needed to to gain greater insight into experimental observations and extrapolate processing outcomes, reducing the need for trial-and-error experimentation.
\par

Models have been developed to understand the nucleation and growth of oxides during ladle deoxidation \cite{ZhangCollisions, lindborgTorssell, turkdogan1967deoxidation, turpin1966nucleation}, arc welding \cite{hongFluid, babu95, hongTTT, kluken1989mechanisms}, laser-directed energy deposition \cite{Eo}, and PBF-LB \cite{Eo}. Typically, these models apply the theories introduced by Kampmann and Wagner \cite{kampmann1970theory, wagner1961theorie} and Langer and Schwartz \cite{langer1980kinetics}, which provide a set of rate equations to describe the nucleation and diffusional growth of spherical precipitates in metastable fluids. The kinetic calculations are informed by thermodynamic models, which describe the driving force for oxide nucleation and growth. While models have been developed to describe the nucleation, growth, and coarsening of Y$_2$O$_3$-based dispersoids in the solid state \cite{nellis2022kinetic, barnard, cunningham2016nano}, these frameworks cannot describe oxide evolution in the liquid melt pool. On the other hand, the models developed to describe oxide formation during arc welding and fusion-based AM processes did not consider oxide phases relevant to ODS alloys. For example, Kluken and Grong studied the formation of Al-Ti-Si-Mn inclusions in steel welds \cite{kluken1989mechanisms}. Babu \textit{et al.} investigated a variety of oxides in low alloy steel welds, including Al$_2$O$_3$ and SiO$_2$ \cite{babu95}. Eo \textit{et al.} focused on the formation of nanometric Si-Mn-Cr-O dispersoids in 316 L stainless steel fabricated with laser-directed energy deposition and PBF-LB \cite{Eo}. Unlike the current investigation, these studies \cite{Eo, babu95} obtained temperature histories from heat transfer simulations, rather than coupling the models to experimental melt pool temperature measurements. Further, the prior work did not probe the upper limit of the dispersoid number density that can be achieved with PBF-LB, nor did it leverage the models to predict trends in oxide size and number density across the processing space, which are key contributions of the current investigation. \par

In this work, the feasibility of PBF-LB as an alternative manufacturing route for ODS alloys is assessed through a combination of modeling and experimentation. It is hypothesized that shorter solidification times (low laser power, high scanning velocity) will yield smaller dispersoid sizes and higher number densities. It is unknown, however, whether solidification is sufficiently rapid to achieve comparable number densities to powder metallurgy. The results provide insight into the following questions: (1) To what extent do the dispersoid size and number density depend on laser power and scanning velocity (which influence solidification time)? (2) What are the limits of the dispersoid size and number density that can be achieved with PBF-LB? (3) What experimental strategies can limit the dispersoid size and increase number density? \par

To address these questions, bulk samples of a Ni-20Cr $+$ 1 wt.\% Y$_2$O$_3$ ODS alloy are fabricated using a variety of laser power and scanning velocity combinations, and a modeling framework is adapted from the welding literature \cite{hongFluid, babu95, hongTTT} to understand the evolution of the oxide nanoparticles during processing. These details are discussed in Section 2. Section 3 presents microstructural observations at a variety of length scales, which are compared against model predictions to reveal the stages of oxide evolution during processing. Finally, in Section 4, limitations of the model framework are explained and calculations are performed to establish the time scales for collision coarsening of dispersoids in the melt pool. Though dispersoid collisions constrain the final number density that can feasibly be achieved, model-informed experimental strategies are proposed to access higher number densities than achieved in the current investigation.

\section{Methods}
\subsection{Experimental}
\subsubsection{Feedstock powder}
The base Ni-20Cr alloy powder (15 $\mu$m $-$ 45 $\mu$m, Powder Alloy Corporation) was coated with 1 wt.\% Y$_2$O$_3$ nanoparticles (100 nm $-$ 200 nm, American Elements) using an acoustic mixing process \cite{nasa20}. The composition of the coated feedstock was measured with inductively coupled plasma optical emission spectroscopy (ICP-OES), combustion-infrared detection, and inert gas fusion (Table \ref{tab:powderComp}). \par

\begin{figure}[bh]
    \centering
    \includegraphics[width=0.3\linewidth]{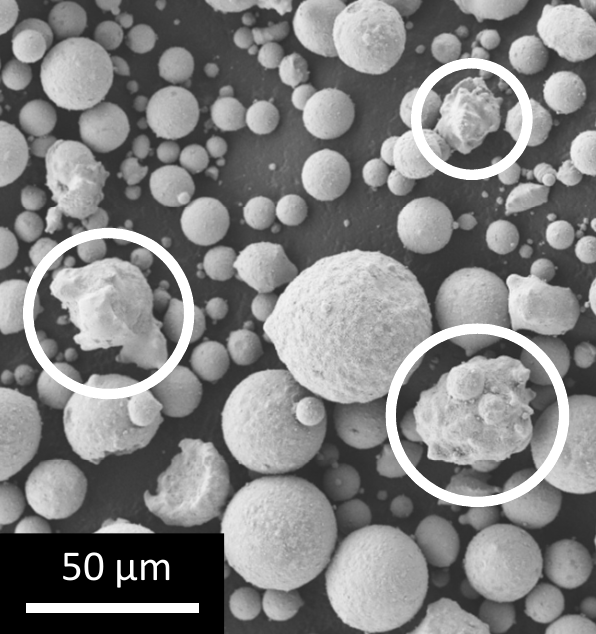}
    \caption{Scanning electron microscopy image of Ni-20Cr (15 $\mu$m $-$ 45 $\mu$m) coated with 1 wt.\% Y$_2$O$_3$ nanoparticles (100 nm $-$ 200 nm). Some of the powder particles have an irregular morphology (circled in white), which affects the uniformity of the oxide coating.}
    \label{fig:feedstock}
\end{figure}

The acoustic mixing process results in a dense Y$_2$O$_3$ coating and generally preserves the sphericity of the powder (Fig.~\ref{fig:feedstock}), which provides good flowability \cite{nasa20} and enables high packing density in the powder bed. Though the coating process promotes good adhesion between the alloy powder and the oxide nanoparticles, the coating is not entirely homogeneous due to the irregular morphology of certain powder particles (Fig. \ref{fig:feedstock}).\par

While Ni-20Cr is not commonly used in industrial applications, the use of a binary alloy simplifies thermodynamic modeling and isolates the effect of laser power and scanning velocity on the as-printed dispersoid characteristics. In multi-component commercial alloys, elements such as Si \cite{zhong, ghayoor}, Al \cite{kenel, boegelein, hunt}, and Ti \cite{horn} can interact with the added Y$_2$O$_3$, which complicates the evolution of the dispersoids during processing.

\begin{table}[bh]
    \centering
    \small
    \caption{Composition of feedstock powder measured with ICP-OES, combustion-infrared detection, and inert gas fusion at EAG Laboratories (Liverpool, NY).}
    \label{tab:powderComp}
    \begin{tabular}{ccccccc}
        \toprule
        \textbf{Element}&Ni&Cr&Y&O&N&S \\
        \midrule
        \textbf{wt.\%}&79.4&19.7&0.85&0.27&280 ppm&11 ppm\\
        \bottomrule
    \end{tabular}
\end{table}

\begin{table}[th]
    \centering
    \small
    \caption{Experimental laser power and scanning velocity combinations. Single tracks and STEM characterization were only performed for Samples 1, 3, 4, and 6.}
    \label{tab:PVcombos}
    \begin{tabular}{ccccccc}
        \toprule
        \textbf{Sample No.}&\textbf{1}& \textbf{2}&\textbf{3}&\textbf{4}&\textbf{5}&\textbf{6} \\
        \midrule
        Laser power (W)&200&300&300&300&300&200 \\
        Scanning velocity (mm/s)&200&400&800&1200&1600&900 \\
        Volumetric energy density (J/mm$^3$)&317&238&119&79.4&59.5&70.5\\
        \bottomrule
    \end{tabular}
\end{table}

\subsubsection{Printing samples with PBF-LB}
Six bulk samples (8 $\times$ 8 $\times$ 5.5 mm$^3$) were fabricated on a stainless steel build plate with the Trumpf TruPrint 3000 PBF-LB system using the laser power and scanning velocity combinations in Table \ref{tab:PVcombos}. A schematic of the build plate is provided in \ref{sec:buildSetup}. In Table \ref{tab:PVcombos}, the volumetric energy density ($E$) is calculated as $E=\frac{P}{VHL}$, where $P$ is the laser power, $V$ is the scanning velocity, $H$ is the hatch spacing, and $L$ is the layer height. Previous work indicated that oxide-coated powder requires higher energy densities than uncoated powder to achieve full melting \cite{nasa20, zhong, spierings}, so the experimental laser power and scanning velocity combinations favor higher energy densities than suggested by empirical processing windows for Alloy 718 \cite{agrawal} and Alloy 625 \cite{criales2017laser}. Though these commercial alloys have a different composition from the Ni-20Cr feedstock, they serve as a useful point of reference for the selection of processing conditions since Ni-20Cr has not been studied extensively in the AM literature. A wide range of scanning velocities (200 mm/s $-$ 1600 mm/s) were selected to probe the effect of solidification time on the dispersoid characteristics in the as-fabricated samples (where higher scanning velocity generally yields a shorter solidification time). The scanning pattern was rotated by 67 degrees between layers. The oxygen content in the build chamber was maintained at less than 0.1\% to minimize the absorption of atmospheric oxygen into the melt pool. Other machine settings were held constant for all samples and are provided in Table \ref{tab:machineSettings}. \par

\begin{table}[htbp]
    \centering
    \small
    \caption{PBF-LB machine settings.}
    \label{tab:machineSettings}
    \begin{tabular}{cc}
        \toprule
        \textbf{Parameter}&\textbf{Value} \\
        \midrule
        Hatch spacing&70 $\mu$m\\
        Layer height&45 $\mu$m\\
        Spot size&100 $\mu$m\\
        Preheat temperature&25 ℃\\
        \bottomrule
    \end{tabular}
\end{table}

\subsubsection{Melt pool surface temperature measurement} \label{surfTempMethods}
The evolution of the oxide nanoparticles depends on the temperature history they experience in the melt pool. Peak temperatures in the melt pool cannot be adequately predicted by conduction-based models due to their inability to capture heat transfer out of the laser interaction region via Marangoni flows and alloy vaporization \cite{stump}. Therefore, melt pool surface temperatures were measured experimentally using the two-color single-camera technique \cite{myers}. \par

During fabrication of the bulk samples, a thin pad (20 layers in thickness) was fabricated using a laser power of 300 W and a scanning velocity of 800 mm/s. After the first 20 layers were completed, four single track scans (5 mm length, spaced 0.7 mm apart) were performed atop the pad using the laser power and scanning velocity combinations for Samples 1, 3, 4, and 6. The single tracks were performed atop the thin pad of the ODS alloy to reduce the influence of the stainless steel build plate, which has a different composition from the ODS alloy. Videos of the single track scans were recorded with a Photron FASTCAM mini-AX200 color camera at 22,500 frames/s and a field of view of 512 $\times$ 512 pixels (2.8 $\times$ 2.8 mm$^2$). The small field of view constrains the number of single tracks that can be imaged with the camera (\ref{sec:buildSetup}). Therefore, the laser power and scanning velocity combinations for Samples 1, 3, 4, and 6 were used to obtain a representative sample of the processing space, capturing two different laser powers and four different scanning velocities. The camera and a magnification lens (Model K2 DistaMax™ Long-Distance, Infinity USA) were mounted atop the front view port of the TruPrint machine. A tri-band filter (Edmunds Optics 87-246) was placed in the magnification lens and the view port glass was replaced with two 750 nm dielectric short pass filters (Edmunds Optics Hot Mirror 64-460) to block reflected laser light. \par

Each single track was scanned multiple times to record videos at seven different exposure times (0.26 $\mu$s $-$ 98.29 $\mu$s). The exposure time (i.e. integration time) of the camera determines how much radiant energy is incident on the camera’s sensor given the radiance of the scene \cite{myers}. The radiance, or brightness, of an emissive body increases with temperature. The camera sensor has both a noise floor and saturation limit. In order for the temperature to be accurately measured, the pixel signal must be between the noise floor and saturation limit. Thus, the lower the temperature, the higher the exposure time must be for the signal to surpass the noise floor of the sensor. Similarly, the higher the temperature, the lower the exposure time must be for the signal to be below the saturation limit of the sensor. Thus, shorter exposure times enable measurement of high temperatures ($<$4000 K) near the laser spot and longer exposure times enable measurement of lower temperatures ($>$2000 K) near the edge of the melt pool. The temperature measurements for several exposure times are then combined to build a composite temperature field of the top surface of the melt pool. Not all of the recorded videos were needed to construct the composite temperature fields (e.g., due to excessive saturation, plume interference, or overlap with other videos), so a subset of five exposure times (0.66 $\mu$s, 1.99 $\mu$s, 5 $\mu$s, 20 $\mu$s, and 42.73 $\mu$s) was used. To build the composite fields, ten frames from each exposure time are averaged together (50 frames total), neglecting pixels where data from less than five frames exists. The construction of the composite temperature field is demonstrated in \ref{sec:compositeConstruction}. From the composite images, a centerline temperature profile is extracted for each of the four laser power and scanning velocity combinations. \par

There are three limitations of the two-color approach to measure melt pool surface temperatures. First, the recorded videos are affected by plume interference, causing a high-temperature, low signal region to appear at the front of the melt pool \cite{myers}. To help prevent this region from impacting the analysis, pixels with a high temperature but low irradiance are removed during the construction of the composite temperature fields. It is important to note that plume artifacts cannot be eliminated entirely by post-process image filtering, which introduces uncertainty to the measured temperatures, especially in the laser interaction region and next to the melt pool, in the direction of the argon flow. Second, since each single track is scanned multiple times (once for each exposure time), the topology of the printing surface changes as successive tracks are deposited. This topological change introduces uncertainty to the time-averaged composite temperature fields. Nevertheless, since the modeling framework aims to predict the average evolution of an ensemble of oxide particles in the melt pool, this uncertainty is not expected to significantly affect the predicted trends in dispersoid size and number density. Third, this two-color setup cannot sense temperatures near the alloy melting point without significant motion blurring and plume obstruction, which is discussed in Section \ref{sec:surfTempResults}.

\subsubsection{Microstructural characterization}
After printing, the bulk samples were removed from the build plate with electrical discharge machining and sectioned in half to reveal the $x$-$z$ cross section of each sample (where +$x$ is parallel to the surface of the build plate and +$z$ is parallel to the build direction). Samples were first mounted in a conductive medium, then ground with 240-grit SiC until plane. Grinding was followed by polishing with 9 $\mu$m diamond suspension, followed by 3 $\mu$m diamond suspension. Colloidal silica (20 nm $-$ 60 nm) was used for the final polishing step. \par

Micron-scale oxide inclusions were observed with a Quanta 600 scanning electron microscope (SEM) equipped with an energy dispersive spectroscopy (EDS) sensor at Carnegie Mellon University. Nanometric oxide dispersoids were observed with a Titan Themis 300 scanning transmission electron microscope (STEM) at Texas A\&M University. Lamellae were extracted from the center of the $x$-$z$ cross-section, cleaned with 5 kV Ga+ ions, and thinned with 30kV Ga+ ions using a Tescan LYRA-3 Model GMH focused ion beam microscope. STEM images were acquired at 300 kV. Since melt pool surface temperatures were not measured for Samples 2 or 5 due to experimental constraints, centerline temperature profiles are not available for those samples. Therefore, model calculations and STEM characterization were performed only for Samples 1, 3, 4, and 6. \par

Electron energy loss spectroscopy (EELS) was performed on the Sample 4 lamella to obtain a representative measurement of the film thickness. EELS measurements were acquired at depths between 0 $\mu$m $-$ 7 $\mu$m with step size 1 $\mu$m, and these measurements were repeated seven times to reduce uncertainty (49 total measurements). The measured mean film thickness is 103 nm (with a standard deviation of 3 nm). Therefore, the number densities of dispersoids in the PBF-LB samples were calculated assuming a uniform film thickness of 103 nm. Micron-scale inclusions were measured with image segmentation in Python and nanometric dispersoids were measured by hand with ImageJ (version 1.53t) \cite{imageJ}. Stereological corrections to the measured particle size distributions were applied using the stereology module of GrainSizeTools \cite{grainsizetools}.

\subsection{Modeling framework}

\subsubsection{Model assumptions}\label{assumptions}
The time- and temperature-dependent evolution of the nanometric Y$_2$O$_3$ particles in liquid Ni-20Cr is modeled following the approach developed by Babu \textit{et al.} to understand the formation of micron-scale oxide inclusions during arc welding processes \cite{hongFluid, babu95, hongTTT}. Before describing the calculation procedure, it is important to first clarify the objective of the modeling approach. The current framework is designed to interpolate the dispersoid size distribution within the experimental processing space and extrapolate the outcome of new processing strategies. The objective of this investigation is not to construct a model with the highest possible fidelity, accounting for all of the complex physics of the PBF-LB process. To make this model computationally-tractable, some simplifying assumptions must be made. Many of these assumptions were made in prior modeling works \cite{Eo, hongFluid, babu95, hongTTT}, but the impact of the assumptions was not discussed in detail. Therefore, extra care is taken to understand the consequences of these assumptions in the current work. \par

(1) Stoichiometric Y$_2$O$_3$ is the only oxide phase considered. When many deoxidizing elements are present, oxides with the highest thermodynamic stability are expected to form first \cite{babu95}. Since Y$_2$O$_3$ has a much higher stability than Ni-oxide and Cr-oxide (\ref{sec:ellingham}), any dissolved O in the liquid alloy is likely to be consumed by Y, limiting the formation of oxides other than Y$_2$O$_3$. Therefore, the impact of this assumption is expected to be negligible. \par

(2) The melting and vaporization of Y$_2$O$_3$ during processing is not considered. Accounting for these physics would require a high-fidelity process model including heat transfer and fluid dynamics, which is beyond the scope of the current investigation. The equilibrium vaporization temperature of Y$_2$O$_3$ exceeds 4500 K, which is much higher than the peak temperatures observed experimentally. Therefore, oxide vaporization is expected to be negligible. On the other hand, the equilibrium melting temperature of Y$_2$O$_3$ is 2700 K, so that melting of the oxide is likely to occur during PBF-LB. It is hypothesized that melting of the oxide during processing will accelerate the time scales for oxide coarsening. Upon melting, deposits of the liquid oxide can deform under the influence of turbulent melt pool flows. This deformation increases the oxide’s surface area beyond the assumed spherical shape, increasing the probability of particle collisions. Therefore, the assumption that the Y$_2$O$_3$ nanoparticles remain solid and spherical represents the best-case scenario with respect to predicting the upper limits of the dispersoid number densities achievable with PBF-LB. \par

(3) Microsegregation is neglected. Upon the onset of solidification, dissolved Y and O atoms will be rejected from the solid into the liquid, increasing their concentration near the solidification front. This effect is important for oxide phases with comparatively low thermodynamic stability (e.g., Si-Mn-Cr-O \cite{Eo}), which require a higher supersaturation of dissolved solute to drive oxide nucleation. However, for highly thermodynamically-stable oxides like Y$_2$O$_3$, it is expected for oxide nucleation to occur at more modest solute supersaturations, which decreases the impact of microsegregation. In alignment with this expectation, the solubility limit of Y$_2$O$_3$ in liquid Ni-20Cr calculated from the modeling framework is negligible ($<$0.01 wt.\%) near the melting temperature of the alloy (1690 K). This result implies that nearly all of the dissolved Y$_2$O$_3$ precipitates out of solution prior to the onset of solidification, which limits the influence of microsegregation. \par

(4) Local equilibrium is assumed at the interface between the oxide and the liquid alloy (i.e., the Gibbs energy of the reactants is equal to that of the products for the deoxidation reaction) \cite{hongTTT}. Since the maximum oxide growth rate predicted by the model framework is less than 1 mm/s (whereas meaningful deviations from equilibrium typically occur at greater than 1 m/s \cite{aziz1982model}), the impact of this assumption is expected to be negligible. \par

(5) All particles in the system experience the same bulk (i.e., far-field) Y and O concentration at a given instant in time \cite{babu95}. This assumption is an approximation of the conditions in the melt pool, where some oxide particles will be more closely spaced than others, causing overlap in their local solute concentration fields and an accompanying reduction in the driving force for both diffusional dissolution and growth. Therefore, the model will overpredict the rate of oxide growth/dissolution compared to the experimental conditions. This point is further discussed in Section \ref{sec:modelResults}. \par

(6) Growth and dissolution of the oxide is limited by the diffusion of O \cite{Eo, babu95}. This assumption would be most-appropriate for a solid-state precipitation process, since the diffusivity of O in solid Ni is much higher than that of Y. However, under melt pool conditions, it is likely that neither the diffusion of Y nor O would be rate-limiting. Three different growth/dissolution mechanisms (O-controlled, Y-controlled, and mixed-controlled) were considered in the model framework, but the model predictions for mean diameter and number density were not significantly affected by the choice of different mechanisms. Quantitative results are presented in \ref{sec:analyzeAssumptions} (Fig. \ref{fig:assumptionComparison}). \par

(7) At a given instant in time, all oxide particles experience the same temperature. It is prohibitively computationally-expensive to explicitly calculate the trajectories of all individual oxide particles in the melt pool \cite{buendiaModel}. Therefore, the temperature profile along the centerline of the melt pool surface is used as an approximation of the average temperature history experienced by the full ensemble of oxide particles. This approximation is acceptable for understanding trends in the dispersoid characteristics across the processing space, which aligns with the primary objectives of the current work. \par

(8) The effects of heat accumulation in the bulk samples are neglected. The temperature histories input to the modeling framework are extracted from single tracks, which experience different heat transfer conditions than the bulk samples. While a thermal camera could be used to monitor the deposition temperature in each layer and then run single tracks at different preheat temperatures, there could also be local track-level variations within a layer. In addition to these trackwise and layerwise temperature variations, remelting between layers could influence the oxide evolution. To balance the practicality of the problem scope and the scientific contribution, the thermal images of the single track melt pools are used as an approximation of the melt pool conditions in the bulk samples. In Section \ref{sec:modelResults}, the trends in the dispersoid sizes and number densities observed in the bulk samples agree with the trends predicted by the modeling framework. Therefore, heat accumulation in the bulk samples is not expected to affect the conclusions of the current work.

(9) Inefficient collisions between oxide particles are not considered \cite{babu95}. In other words, when two particles come into contact, they instantly merge together with zero probability of breaking apart. The consequences of oxides breaking apart under the influence of turbulent melt pool flows are discussed in Section \ref{inclusionDiscussion}.\par

(10) Pushing of oxide particles away from the advancing solidification front is neglected. It is known from the casting literature that oxides are pushed ahead of the solidification front by hydrodynamic forces \cite{malmberg2010observed}, extending their residence time in the melt pool. Since longer residence times imply greater coarsening, ignoring this effect represents the best-case scenario with respect to predicting the upper limits of the dispersoid number densities achievable with PBF-LB. \par

\vspace{10px}
\begin{figure}[h]
    \centering
    \includegraphics[width=1\linewidth]{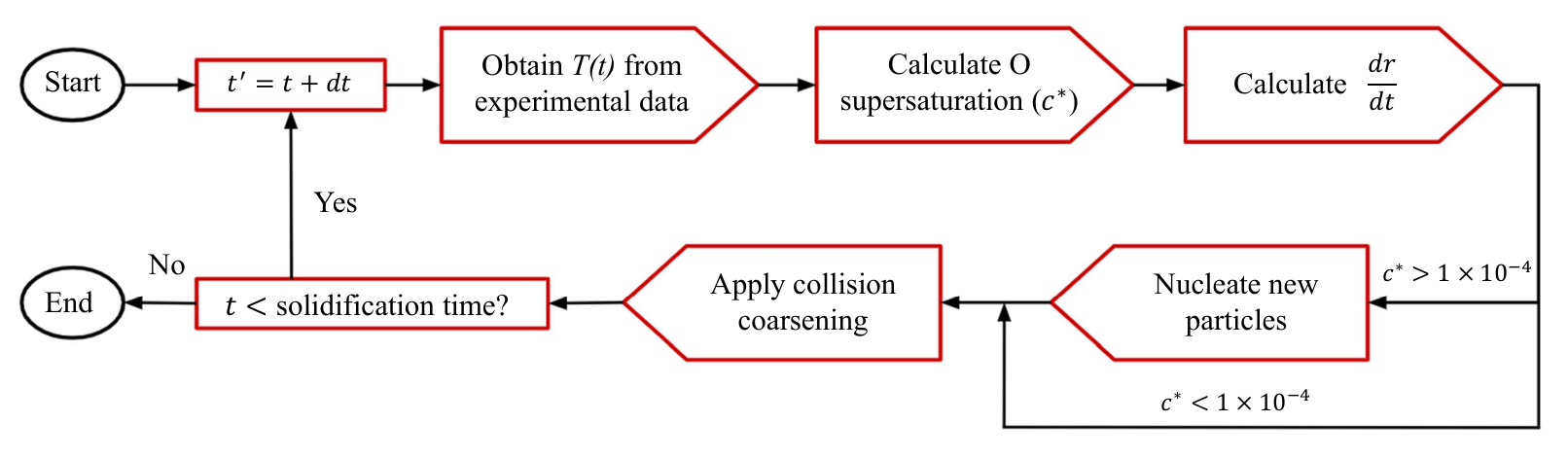}
    \caption{Dispersoid evolution model flowchart.}
    \label{fig:flowchart}
\end{figure}

\subsubsection{Model setup}
The modeling approach is summarized in Fig. \ref{fig:flowchart}. An initial particle size distribution is defined, consisting of an array of particle radii and a corresponding array of particle counts. Unless otherwise specified, the initial distribution consists of 50,000 particles normally distributed 100 nm $-$ 200 nm in diameter. The mean of the initial distribution is 150 nm, with standard deviation 75 nm. The distribution is truncated such that all particles fall between 100 nm $-$ 200 nm in diameter, in accordance with the size distribution reported by the manufacturer of the Y$_2$O$_3$ nanoparticles. Temperature is calculated at each time step based on a polynomial fitted to the measured melt pool centerline temperature profile, where length is converted to time by dividing by the laser scanning velocity.\par

\subsubsection{Oxide dissolution and growth}
Consider the reaction:

\begin{equation}
\label{eq:reaction}
    2\underline{\text{Y}} + 3\underline{\text{O}} \rightleftharpoons \text{Y}_2\text{O}_3 \text{ (s)},
\end{equation}

\noindent where the underlined elements are dissolved in liquid Ni-20Cr. The equilibrium constant for the reaction is \cite{hongTTT}:

\begin{equation}
\label{eq:eqbm}
    k_{eq} = \frac{a_{Y_2O_3}}{a_Y^2a_O^3} = \frac{1}{(f_Yc^i_Y)^2(f_Oc_O^i)^3} = \exp{(-\Delta G^0/RT)},  \\[1.5ex]
\end{equation}

\noindent where $a$ are activities, $f$ are activity coefficients, $c^i$ are solute concentrations at the reaction interface, and $\Delta G^0$ is the standard free energy change for the deoxidation reaction  in Eq. \eqref{eq:reaction}. The solid oxide is taken to be a pure stoichiometric phase, so its activity is set to unity. The activity coefficients for Y and O can be estimated from their empirical first-order interaction parameters ($e$) as:
\begin{align}
\label{eq:activityY}
    \log{f_Y} &= c_Ye_{YY} + c_Oe_{YO} \\
\label{eq:activityO}
    \log{f_O} &= c_Oe_{OO} + c_Ye_{OY}.
\end{align}

\begin{table}[th]
    \centering
    \small
    \caption{Model parameters. Values are taken from the ThermoCalc TCNI12 database \cite{andersson2002thermo} unless otherwise indicated. Note that ThermoCalc predicts the freezing range of Ni-20Cr to be negligible, so separate solidus and liquidus temperatures are not listed.}
    \label{tab:modelParams}
    \begin{tabular}{cccc}
        \toprule
        \textbf{Parameter}&\textbf{Unit}&\textbf{Value}&\textbf{Reference} \\
        \midrule
        Ni-20Cr melting temperature&K&$1690$&$-$ \\
        Ni-20Cr density&kg/m$^3$&$8500$ (solid); $7100$ (liquid)&$-$ \\
        Y$_2$O$_3$ density&kg/m$^3$&$5010$&\cite{y2o3density} \\
        $\Delta G^0$&J/mol&$296T-1.28*10^6$&\cite{turkdogan1980physical} \\
        $e_{YY}$&$-$&$22.3/T-0.006$&\cite{ting1985} \\
        $e_{OO}$&$-$&$-1750/T+0.734$&\cite{sigworth} \\
        $e_{YO}$&$-$&$-2.5$&\cite{ishii} \\
        $e_{OY}$&$-$&$-0.46$&\cite{ishii} \\
        $D_Y$&m$^2$/s&$(5*10^{-12})T-6*10^{-9}$&$-$ \\
        $D_O$&m$^2$/s&$(1.05*10^{-6})\exp({-72500/RT})$&\cite{Odiffusivity} \\
        grad $v$&/s&$5000$&\cite{Eo, ChenPores} \\
        $\mu$&Pa$\cdot$s&$\exp{(2.9*10^{-7}T^2-2.3*10^{-3}T-2.23)}$&$-$ \\
        \bottomrule
    \end{tabular}
\end{table}

\noindent Since the concentration of oxide in the alloy is relatively dilute (1 wt.\%), the activity coefficients do not have a significant impact on the results, as shown in \ref{sec:analyzeAssumptions} (Fig. \ref{fig:assumptionComparison}).\par

Based on the relative fluxes of Y and O for the reaction in Eq. \eqref{eq:reaction}, the mass balance can be written as \cite{hongref3}:

\begin{equation}
\label{eq:stoi}
    c_Y^i = c_Y^b - \frac{2m_Y}{3m_O}\sqrt{\frac{D_O}{D_Y}}(c_O^b - c_O^i),  \\[1.5ex]
\end{equation}
\noindent where $m$ are atomic weights, $D$ are diffusion coefficients, and $c^b$ are bulk solute concentrations in the liquid alloy. From Eqs.~\eqref{eq:eqbm} and \eqref{eq:stoi}, the concentrations of Y and O at the reaction interface ($c_Y^i$ and $c_O^i$) can be calculated at each time step.

\noindent Given $c_O^i$, the dimensionless O supersaturation can be calculated from:

\begin{equation}
\label{eq:cstar}
    c^* = \frac{c_O^b-c_O^i}{c_O^p-c_O^i},  \\[1.5ex]
\end{equation}

\noindent where $c^p$ is the concentration of O inside the oxide particle (21 wt.\% for Y$_2$O$_3$). When the O concentration at the reaction interface is larger than the bulk O concentration, $c^*$ is negative and the oxides in the system dissolve according to \cite{whelanDissolve}:

\begin{equation}
\label{eq:dissolve}
    \frac{dr}{dt} = c^*\left(\frac{D_O}{r}+\sqrt{\frac{D_O}{\pi t}}\right),  \\[1.5ex]
\end{equation}

\noindent where $r$ is the particle radius and $t$ is time. When the O concentration at the reaction interface is smaller than the bulk O concentration, $c^*$ is positive, and the oxides in the system grow according to \cite{christianGrow}:

\begin{equation}
\label{eq:grow}
    \frac{dr}{dt} = \sqrt{\frac{D_Oc^*}{2t}}. \\[1.5ex]
\end{equation}

\noindent The quantities required to evaluate the model equations are listed in Table \ref{tab:modelParams}.

\subsubsection{Oxide nucleation}
An explicit calculation of the oxide nucleation rate would require expressions for the Gibbs energy of both Y$_2$O$_3$ and liquid Ni-20Cr, each as a function of Y concentration, O concentration, and temperature \cite{Eo}. Since these models are not readily available in the literature, an approximate scheme is used, where both the nucleation rate ($I_v=10^{26}$ m$^{-3}$ s$^{-1}$) and the critical radius ($r_c$ = 1 nm) are held constant. Nuclei are only added to the system upon cooling when the O supersaturation ($c^*$) exceeds a threshold value. \par

There are two reasons to justify the use of a constant value for $I_v$. First, Babu \textit{et al.} showed that the oxide nucleation rate in arc welding is approximately constant before dropping rapidly by several orders of magnitude as dissolved solute is consumed through nucleation and diffusional growth \cite{babu95}. This behavior is well-represented mathematically by a function that is equal to a constant when $c^*$ is above the threshold value and equal to zero when $c^*$ is below the threshold value. Second, the high thermodynamic stability of Y$_2$O$_3$ and the rapid cooling rates in PBF-LB encourage rapid diffusional growth of the oxide particles immediately after the onset of nucleation. Since dissolved Y and O are consumed rapidly as new nuclei enter the system, the duration of the nucleation event is typically less than 10\% of the overall solidification time. Therefore, the constant value for $I_v$ can be regarded as an average value over this short time interval (typically less than 0.1 ms). The nucleation rate is set to 10$^{26}$ m$^{-3}$ s$^{-1}$ for all of the calculations in this work, which is within one order of magnitude of the peak nucleation rate calculated by Eo \textit{et al.}~for Si-Mn-Cr-O oxides in steel processed with PBF-LB. Since an explicit calculation of the driving force for nucleation is not performed, the critical radius cannot be calculated. Therefore, the critical radius is held constant at 1 nm, which is comparable to calculated values for Y$_2$O$_3$ in liquid Fe \cite{jiang} and for Si-Mn-Cr-O oxide in liquid steel \cite{Eo}. \par

Calculations were performed to evaluate the sensitivity of the model to the assumed values of $I_v$ and $r_c$, as presented in \ref{sec:nucleationSensitivity}. Since the objective of the current investigation is to establish trends in the dispersoid characteristics across the laser power and scanning velocity space $-$ rather than to precisely predict dispersoid size distributions for a given set of processing conditions $-$ the sensitivities to the assumed values of $I_v$ and $r_c$ are considered to be tolerable.\par

Since the driving force for oxide nucleation is not calculated explicitly in the current model, an alternative criterion must be applied to determine whether new particles can enter the system. Fundamentally, nucleation requires a supersaturation of dissolved solute in the liquid alloy. Since $c^*$ (the dimensionless supersaturation of dissolved O) must be calculated at each time step to determine the rate of diffusional growth/dissolution, it is convenient to also use $c^*$ to control nucleation. A positive value of $c^*$ implies an over-saturation of dissolved solute in the liquid alloy and a negative value of $c^*$ implies an under-saturation. Therefore, the $c^*$ threshold for nucleation is set to 1$\times$10$^{-4}$. This value is chosen to be sufficiently large such that nuclei added during one time step do not dissolve instantly during the following time step. The value must also be sufficiently small such that is does not exceed the maximum value of $c^*$ reached upon cooling, which is typically between 1$\times$10$^{-3}$ and 1$\times$10$^{-2}$. However, since cooling rates in PBF-LB are high and $c^*$ increases with decreasing temperature for a given bulk O concentration, the model predictions for mean diameter and number density do not depend strongly on the $c^*$ threshold. Quantitative comparisons are presented in \ref{sec:nucleationSensitivity} (Fig. \ref{fig:rc_cstarSensitivity}).\par

In general, nuclei can form either through homogeneous nucleation or heterogeneous nucleation. Since the current modeling framework assumes a constant nucleation rate for the oxide particles, the distinction between the two modes of nucleation is neglected. Nevertheless, gaining insight into the mechanisms of oxide nucleation in the melt pool is a compelling area of future work. \par

\subsubsection{Collision coarsening}
Coarsening due to collisions between oxide particles is modeled using a Monte Carlo approach introduced by Garcia \textit{et al.} \cite{garcia}. The model incorporates three sources of collisions: Brownian ($B$) motion of the particles [Eq. \eqref{eq:brownian}], particle entrainment in turbulent ($T$) flows [Eq. \eqref{eq:turbulent}], and Stokes ($S$) collisions due to buoyancy [Eq. \eqref{eq:stokes}]. The corresponding kernels are \cite{Eo, ZhangCollisions,lindborgTorssell, babu95}:
\vspace{8px}
\begin{align}
\label{eq:brownian}
    \beta^{B}_{ij} &= \frac{2k_{\text{B}}T}{3\mu}\left(\frac{1}{r_i}+\frac{1}{r_j}\right)(r_i+r_j)\\[1.5ex]
\label{eq:turbulent}
    \beta^{T}_{ij} &= \frac{4}{3}(r_i+r_j)^3\text{grad} \, v \\[1.5ex]
\label{eq:stokes}
    \beta^{S}_{ij} &= \frac{2\pi g}{9\mu}(\rho_{Ni20Cr}-\rho_{Y_2O_3})(r_i+r_j)^3|r_i-r_j|,
\end{align}

\noindent where $\beta_{ij}$ is related to the collision probability between two particles $i$ and $j$, $r$ is the particle radius, $k_\text{B}$ is the Boltzmann constant, $\mu$ is the dynamic viscosity of the liquid alloy, grad $v$ is the spatial velocity gradient in the melt pool, and $g$ is the gravitational constant. No fitting parameters are used in the collision model. The collision rate ($W_{ij}$) between two particles of a given size depends on the sum of the collision kernels:
\vspace{5px}
\begin{equation}
\label{eq:collRate}
    W_{ij}=(\beta^{B}_{ij}+\beta^{T}_{ij}+\beta^{S}_{ij})/V_{coll},
\end{equation}

\noindent where $V_{coll}$ is the collision volume, set by the number density of oxide particles at the current time step. For computational efficiency, the collision volume includes 200 oxide particles, which are randomly sampled from the full distribution for each collision event. A random pair of particles ($i$ and $j$) collides when \cite{garcia}:

\begin{equation}
\label{eq:collCrit}
    R < \frac{W_{ij}}{W_{max}},  \\[1.5ex]
\end{equation}

\noindent where $R$ is a random number between 0 $-$ 1 and $W_{max}$ is the maximum collision rate across all pairs in the sample. Given the collision rate for the colliding pair $ij$, the waiting time between collisions is calculated from \cite{garcia}:
\vspace{8px}
\begin{equation}
\label{eq:tWait}
    \tau = \frac{2}{N(N-1)W_{ij}}  \\[1.5ex]
\end{equation}

\noindent where $N$ is the number of particles in the collision volume. When two particles collide, they are removed from the system and replaced with a new particle of radius $\sqrt[3]{r_i^3 + r_j^3}$. Since collisions are calculated explicitly for only a subset of all particles, the number of particles removed from the overall system is equal to $V_{sys}/V_{coll}$~multiplied by the number of collisions in the 200-particle sample, where $V_{sys}$ is the total volume of the alloy-oxide system.

\section{Results}
\subsection{Experimental}
\subsubsection{Microstructural characterization}
\begin{figure}[bh]
    \centering
    \includegraphics[width=0.7\linewidth]{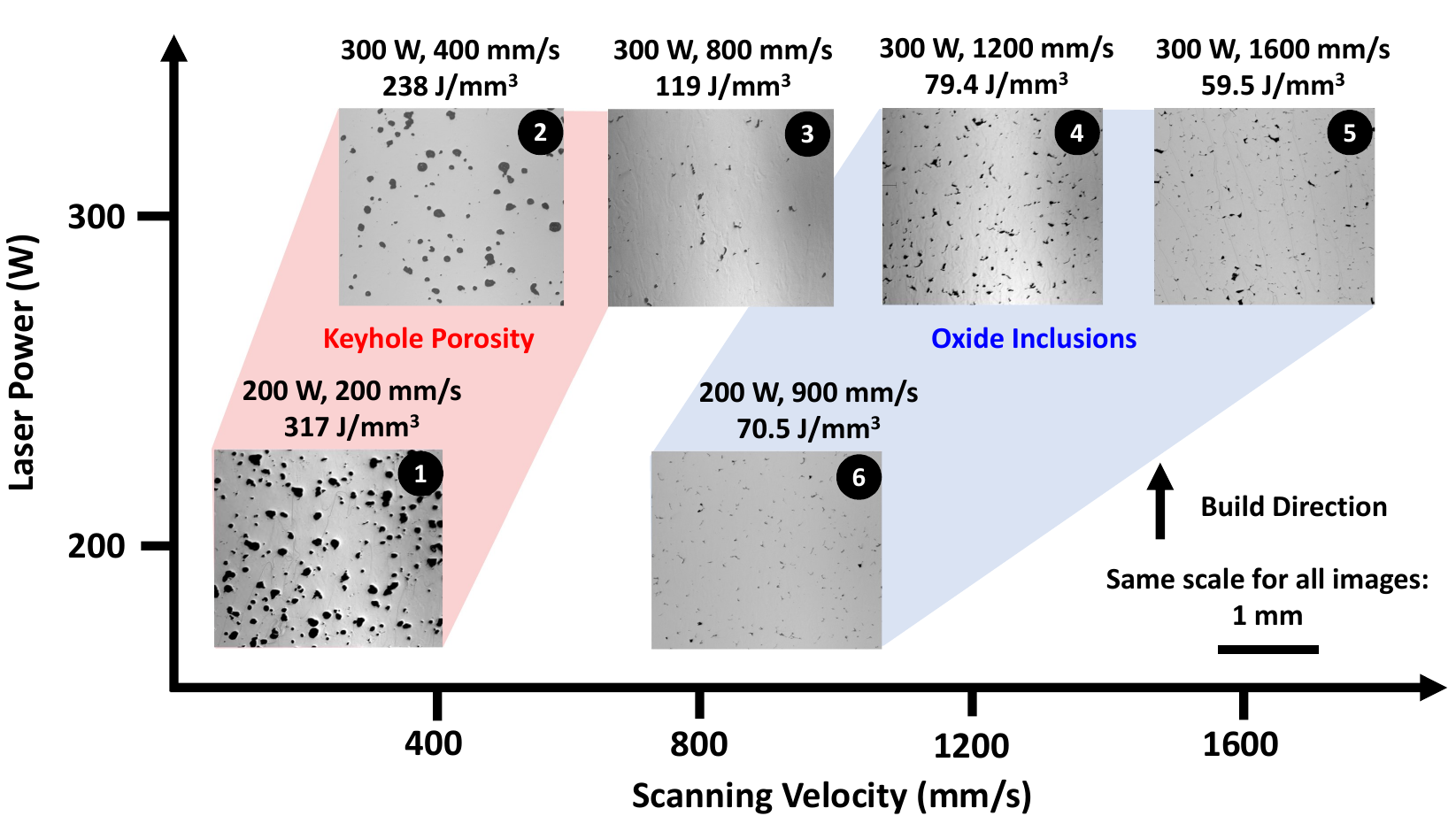}
    \caption{Representative optical micrographs across the experimental laser power and scanning velocity space. Sample numbers are given in the top right corner of each image.}
    \label{fig:opticalPVmap}
\end{figure}

\begin{figure}[th]
    \centering
    \includegraphics[width=0.85\linewidth]{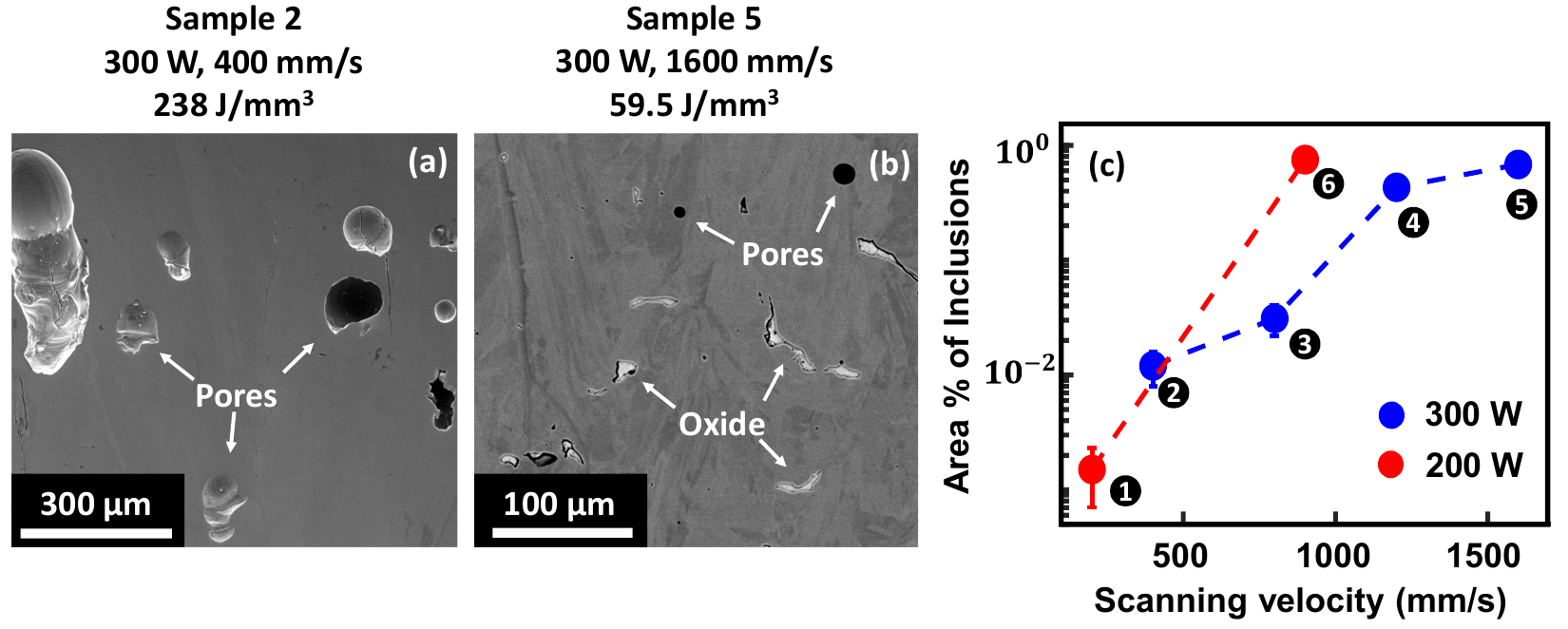}
    \caption{(a) Secondary electron SEM image for Sample 2 (300 W, 400 mm/s); (b) Backscattered electron SEM image for Sample 5 (300 W, 1600 mm/s); (c) Area percentage of irregular micron-scale oxide inclusions. The error bars indicate the standard error in area percentage across 20 different micrographs.}
    \label{fig:UstructureComparison}
\end{figure}

All samples contained micron-scale defects, including pores and oxide inclusions, as shown in Figs.~\ref{fig:opticalPVmap} and \ref{fig:UstructureComparison}. Oxide inclusions can be readily identified in backscattered electron SEM images [Fig. \ref{fig:UstructureComparison}(b)], where they appear brighter than the Ni-20Cr matrix due to Z contrast. The inclusions can be easily distinguished from pores, which appear darker than the matrix in both secondary electron images and backscattered electron images. EDS spectra were collected at a variety of the bright locations identified in the backscattered electron images. An example EDS spectrum is provided in \ref{sec:ellingham} (Fig. \ref{fig:edsSpectrum}). All of the EDS spectra confirm the composition of the oxide inclusions to be consistent with Y$_2$O$_3$, which is expected since Y is a much stronger deoxidizer than the other alloy components (Fig. \ref{fig:ellingham} provides an Ellingham diagram). \par

The area percentages of irregular micron-scale inclusions are plotted in Fig. \ref{fig:UstructureComparison}(c). While the inclusions are almost entirely absent at low scanning velocity, the area percentage nears 1\% with increasing scanning velocity. In the approximation that the inclusions are homogeneously distributed and randomly oriented, the area\% can be equated to the volume\% \cite{Underwood1973}. Under this assumption, it is estimated that as much as 45\% of the Y$_2$O$_3$ added to the feedstock is contained within the micron-scale inclusions, which form at the expense of nanometric dispersoids. It is hypothesized that the inclusions form partly as a consequence of oxide melting, as discussed in Section \ref{inclusionDiscussion}. \par

For Sample 1 ($E$ = 317 J/mm$^3$) and Sample 2 ($E$ = 238 J/mm$^3$), spherical pores (likely the result of keyholing) predominate over oxide inclusions. Cracks were also observed in these samples, commonly found near pores and along the top and bottom surfaces of the samples. In Sample 4 ($E$ = 79.4 J/mm$^3$), Sample 5 ($E$ = 59.5 J/mm$^3$), and Sample 6 ($E$ = 70.5 J/mm$^3$), oxide inclusions predominate over open porosity. Sample 3 ($E$ = 119 J/mm$^3$) contains a mixture of keyhole porosity and oxide inclusions, though the area percentage of inclusions was substantially lower than that in Samples 4, 5, and 6 [Fig. \ref{fig:UstructureComparison}(c)]. Though optimization of the mass density of the printed samples was beyond the scope of the current investigation, more information about the mitigation of defects in ODS alloys fabricated with PBF-LB is available in the literature \cite{kenel, saptarshi, bahadur2023printability}.

\newpage
\begin{figure}[th]
    \centering
    \includegraphics[width=0.9\linewidth]{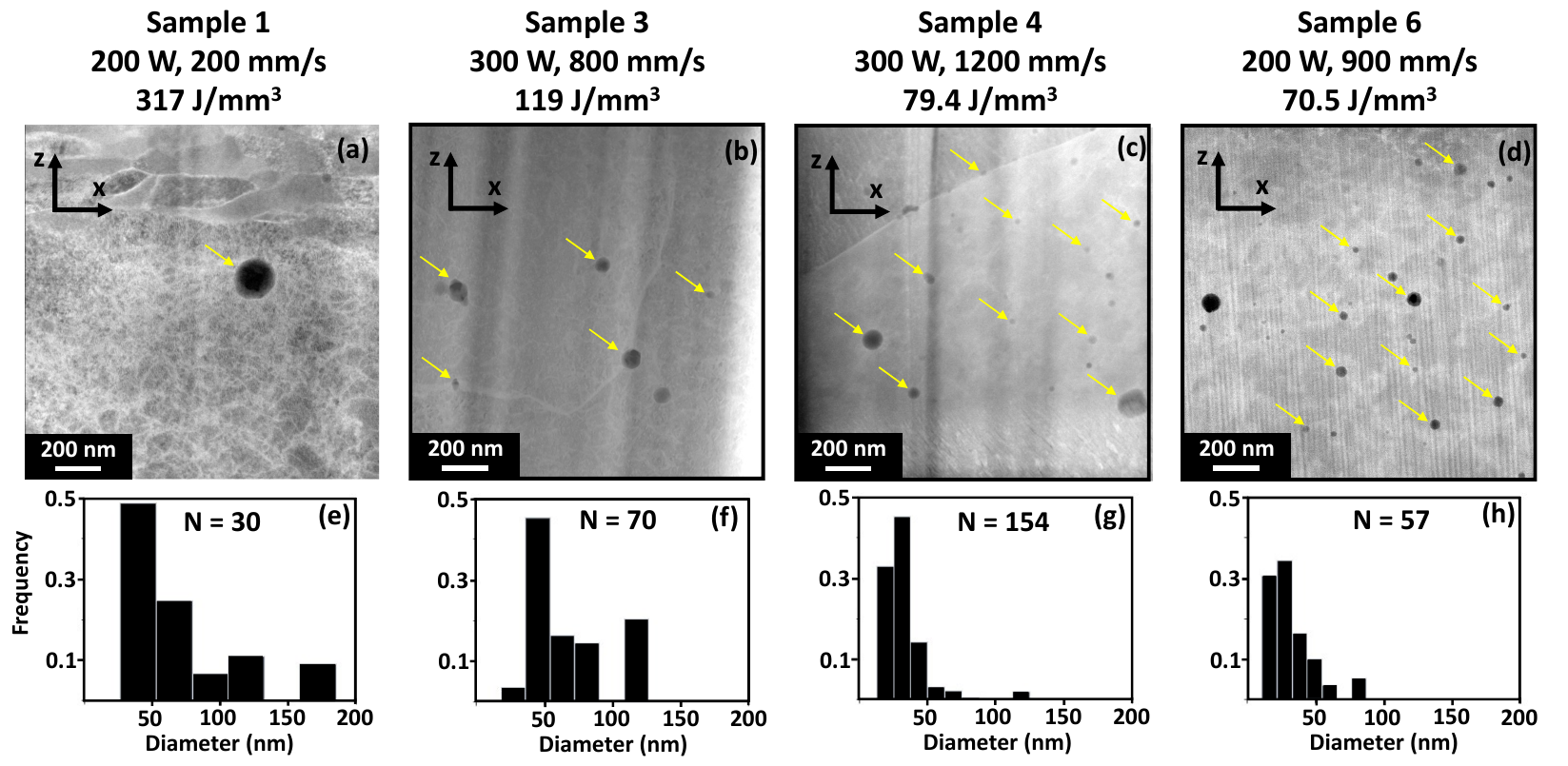}
    \caption{(a)$-$(d) High-angle annular dark field STEM images of oxide dispersoids in bulk microstructure at 63kx magnification. The yellow arrows indicate the locations of dispersoids. To preserve the clarity of the images, some of the dispersoids are not accompanied by yellow arrows. (e)$-$(h) Distribution of dispersoid diameters from multiple micrographs, including stereological correction.}
    \label{fig:psd}
\end{figure}

STEM characterization was performed for Samples 1, 3, 4, and 6, which are the same samples for which melt pool surface temperature measurements were taken. Typical STEM images of the bulk as-fabricated microstructure are shown in Figs.~\ref{fig:psd}(a)$-$\ref{fig:psd}(d). The smallest mean dispersoid diameters were recorded for Samples 4 and 6, taking values of 29 nm and 30 nm. The highest number densities were recorded in the same two samples, both taking a value of $\sim$1$\times$10$^{20}$ m$^{-3}$. The measured dispersoid sizes and number densities are consistent with previous experiments that used similar Ni-based ODS alloys \cite{nasa20, nasa23}. From the micrographs provided in \cite{nasa20, nasa23}, it is estimated that the mean dispersoid diameter was between 33 nm $-$ 68 nm with number density 5$\times$10$^{19}$ m$^{-3}$ $-$ 1$\times$10$^{20}$ m$^{-3}$, similar to Samples 4 and 6 in the current investigation. \par

\subsubsection{Melt pool surface temperature measurements} \label{sec:surfTempResults}
\begin{figure}[bh]
    \centering
    \includegraphics[width=0.7\linewidth]{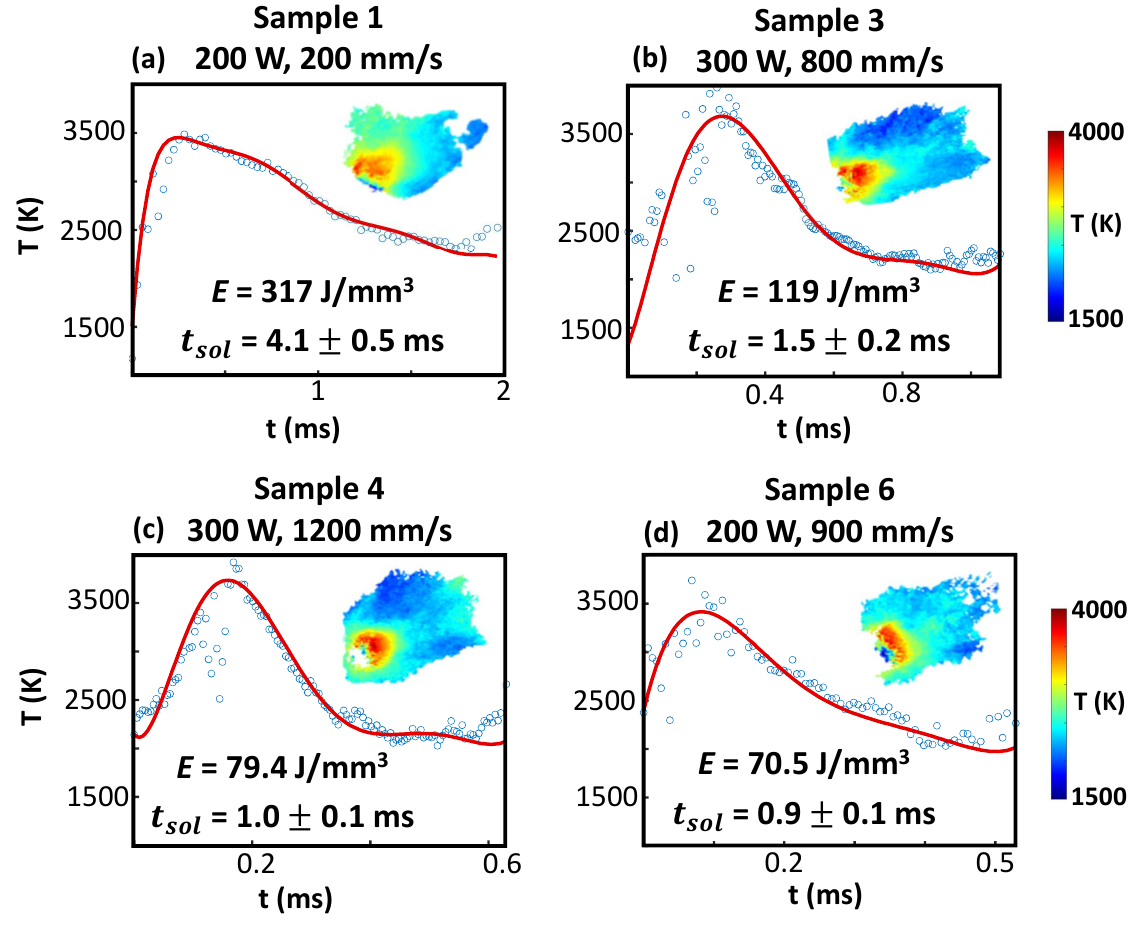}
    \caption{Measured temperatures along the centerline at the top surface of the melt pool and estimated solidification times ($t_{sol}$). The assumed linear cooling period to the alloy melting point is not shown. The inset images are composite temperature fields of the melt pool surface. The gaps in the composite images result from plume interference (discussed in Section \ref{surfTempMethods}).}
    \label{fig:cameraTemps}
\end{figure}

The measured melt pool surface temperatures and fitted polynomials are shown in Figs.~\ref{fig:cameraTemps}(a) $-$ \ref{fig:cameraTemps}(d). The peak temperatures of the liquid alloy ($\sim$3500 K) exceed its equilibrium boiling point (3000 K), which is possible because of the non-equilibrium thermodynamic conditions in the melt pool (e.g., transient heating, recoil pressure, and low accommodation coefficient \cite{myers}). For the range of laser power and scanning velocity studied, the solidification time ($t_{sol}$) is estimated to vary from 0.9 ms $-$ 4.1 ms and decreases with increasing scanning velocity for a given laser power. \par

The estimated solidification times (calculated as melt pool length divided by the scanning velocity) are longer than those obtained from the raw camera measurements, which underestimate melt pool length. The underestimation is a consequence of the limited dynamic range of the camera, which requires long exposure times ($>$100 $\mu$s) to sense temperatures near the alloy melting point (1690 K) \cite{myers}. Therefore, melt pool lengths (and uncertainties) are estimated from single track experiments performed using Alloy 625 substrates \cite{heigel2018measurement}, which is similar in composition to Ni-20Cr. Further details about the estimation procedure are provided in \ref{sec:poolLength}. The extra length is added to the tail of the melt pool and linear cooling is assumed between the final measured temperature and the alloy melting point (1690 K). The linear cooling period does not significantly affect the modeling results, as shown in \ref{sec:analyzeAssumptions}.\par

\subsection{Modeling}\label{sec:modelResults}
\begin{figure}[bh]
    \centering
    \includegraphics[width=0.6\linewidth]{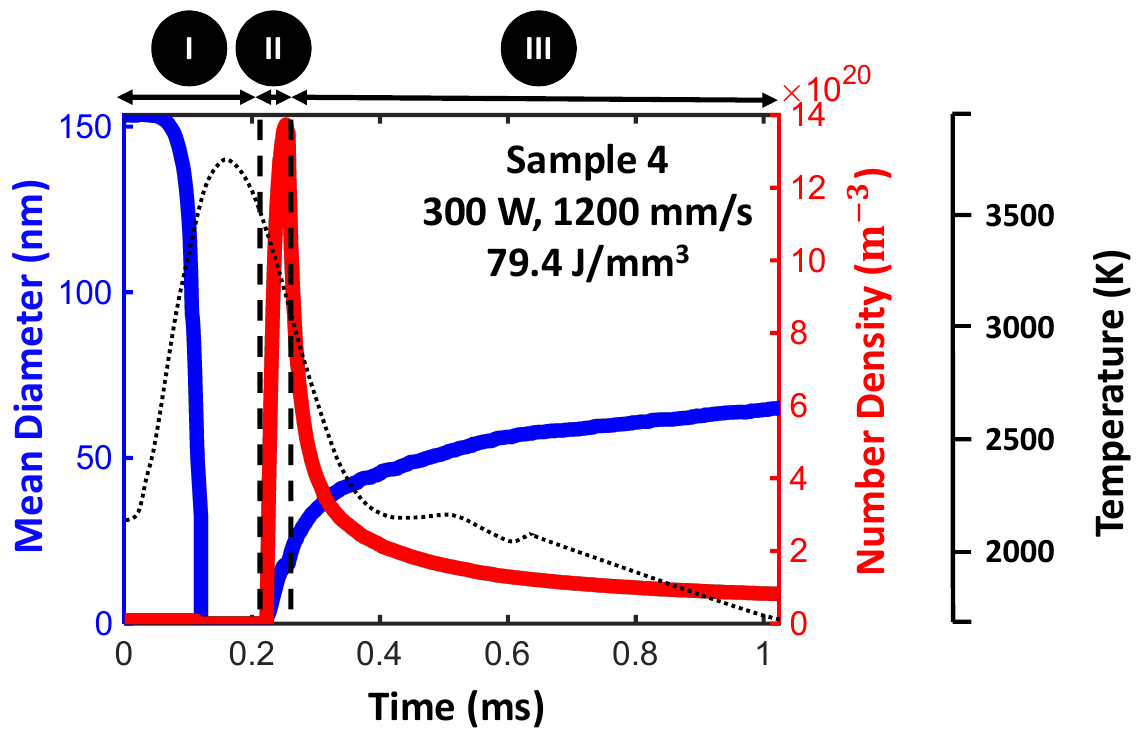}
    \caption{Predicted evolution of the mean diameter and number density of 1 wt.\% Y$_2$O$_3$ nanoparticles for the Sample 4 (300 W, 1200 mm/s) temperature history, where I = dissolution, II = nucleation and diffusional growth, III = growth via collision coarsening.}
    \label{fig:oxideEvolution}
\end{figure}

The predicted time evolution of the mean diameter and number density of the oxide dispersoids for the Sample 4 temperature history [300 W, 1200 mm/s; Fig. \ref{fig:cameraTemps}(c)] is shown in Fig. \ref{fig:oxideEvolution}. The oxide evolution can be divided into three phases: (I) dissolution upon heating; (II) nucleation and diffusional growth upon cooling; and (III) growth via collision coarsening. Due to the high peak temperatures, full dissolution of the feedstock oxides is predicted despite their high thermodynamic stability. Note that the extent of dissolution only becomes significant at temperatures far exceeding the equilibrium melting point of Y$_2$O$_3$ (2700 K), which is discussed in Section \ref{inclusionDiscussion}. Full dissolution leads to a high supersaturation of dissolved O upon cooling, providing a high driving force for nucleation and diffusional growth over a short time interval. Once most of the dissolved solute is consumed, nucleation and diffusional growth become negligible, and collision coarsening begins to control the evolution of the system. At high number density, particles are spaced closely together, which results in a high probability of collisions and rapid coarsening. Each collision results in a net removal of one particle from the system (since two particles combine to form one larger particle), leading to a decrease in number density and an increase in mean diameter. As the number density decreases, the probability of collisions also decreases, and the rate of coarsening slows. \par

\newpage
\begin{figure}[th]
    \centering
    \includegraphics[width=0.9\linewidth]{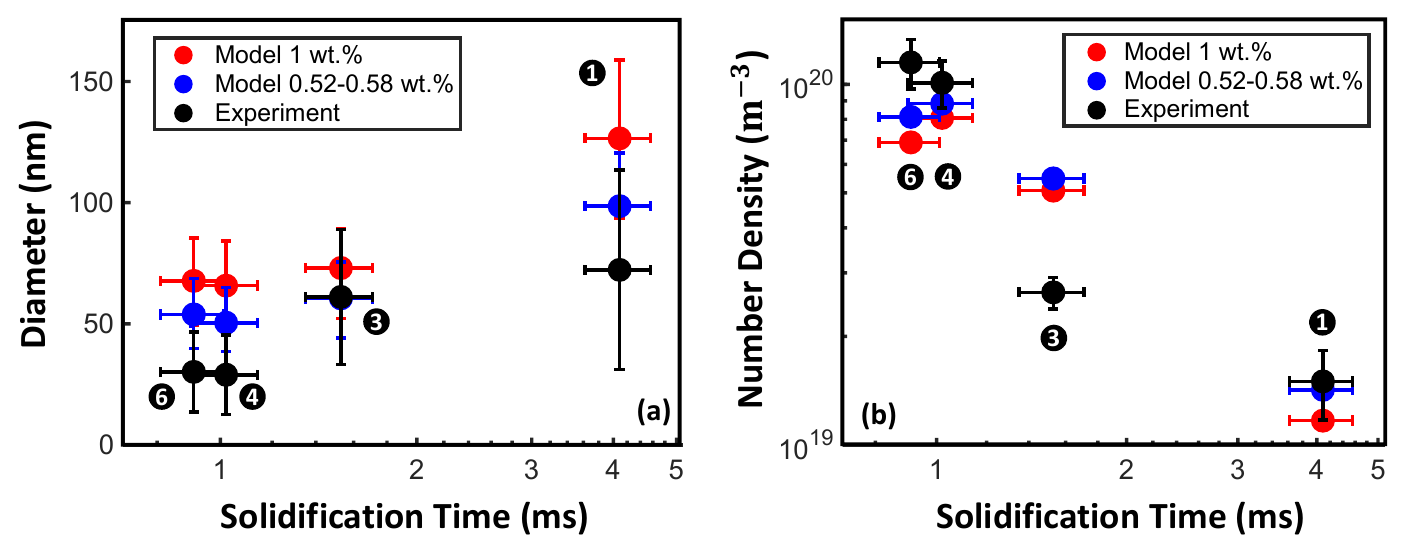}
    \caption{Comparison of the model predictions against the experimental results for Sample 1 (200 W, 200 mm/s), Sample 3 (300 W, 800 mm/s), Sample 4 (300 W, 1200 mm/s), and Sample 6 (200 W, 900 mm/s) using the temperature histories from Fig.~\ref{fig:cameraTemps}. The sample numbers are indicated near the corresponding data points. All horizontal error bars are computed from the uncertainty in the measured melt pool lengths \cite{heigel2018measurement}. Blue datapoints are adjusted for oxide losses due to spatter and agglomeration, as discussed in Section \ref{lossDiscussion}. (a) Comparison of dispersoid diameters. The experimental vertical error bars represent one standard deviation across all measured diameters for a given sample. The model vertical error bars are the 25th and 75th percentiles of the final oxide size distribution. (b) Comparison of dispersoid number densities. The experimental vertical error bars are the standard error in number density across all micrographs for each sample.}
    \label{fig:solTime}
\end{figure}

The model predictions for the dispersoid diameter and number density for Samples 1, 3, 4, and 6 are compared against the STEM observations in Figs.~\ref{fig:solTime}(a) and \ref{fig:solTime}(b). These predictions generally align with experimental trends. Though it is expected that the number density will decrease monotonically with increasing solidification time, the model predicts Sample 6 ($t_{sol}$ = 0.9 ms; 200 W, 900 mm/s) to have a lower number density of dispersoids than Sample 4 ($t_{sol}$ = 1.0 ms; 300 W, 1200 mm/s). From Figs. \ref{fig:cameraTemps}(c) and  \ref{fig:cameraTemps}(d), it is clear that the measured peak temperature is higher for Sample 4 than for Sample 6. Consequently, oxide nucleation occurs comparatively earlier using the Sample 6 temperature history, yielding a longer oxide residence time (defined as the time difference between oxide nucleation and alloy solidification). Since the experimental results imply that number density decreases monotonically with increasing solidification time, the discrepancy in the model predictions is attributed to plume interference, which is more apparent for the Sample 6 temperature history than for Sample 4. \par

By setting the concentration of Y$_2$O$_3$ to the nominal value (1 wt.\%), the model overpredicts the dispersoid diameters for all solidification times. There are three potential reasons for the overprediction of oxide diameters. First is the assumption that all particles in the system experience the same bulk Y and O concentration at a given instant in time, as discussed in Section \ref{assumptions}. This assumption neglects the potential overlap between the solute concentration fields surrounding different particles in the melt pool, causing the model to overpredict the magnitude of $c^*$ (and thus $\frac{dr}{dt}$) during both growth and dissolution.\par

The second reason is the overprediction of the dissolution of Y$_2$O$_3$ particles added to the feedstock. While measured melt pool surface temperatures exceed 3500 K, most of the oxides are expected to be entrained in melt pool flows below the surface, where temperatures are comparatively lower. Within the framework of the computational model, all particles experience the same temperature at a given instant in time, which is not representative of experimental conditions, where each particle may experience a different temperature history based on its individual trajectory in the melt pool. If the peak temperatures experienced by the oxides are lower than those shown in Fig. \ref{fig:cameraTemps}, particles added to the feedstock may only partially dissolve. With partial dissolution, the concentration of dissolved solute in the liquid alloy is diminished, which reduces the subsequent driving force for diffusional growth and limits particle diameters to smaller values. \par

The third reason is the assumption that the total mass of oxide in the system remains constant over time. In reality, a considerable fraction of oxide added to the feedstock is lost to spatter and agglomeration during processing \cite{kenel, haines}. As such, the concentration of nanometric Y$_2$O$_3$ in printed parts is typically much lower than the nominal feedstock concentration. These losses limit the driving force for diffusional growth and limit oxide sizes to smaller values. Adjusted results accounting for oxide losses are explained in Section \ref{lossDiscussion} and are presented in Fig. \ref{fig:solTime}. \par

\section{Discussion}
\subsection{Micron-scale oxide inclusions}\label{inclusionDiscussion}
\begin{figure}[bh]
    \centering
    \includegraphics[width=0.35\linewidth]{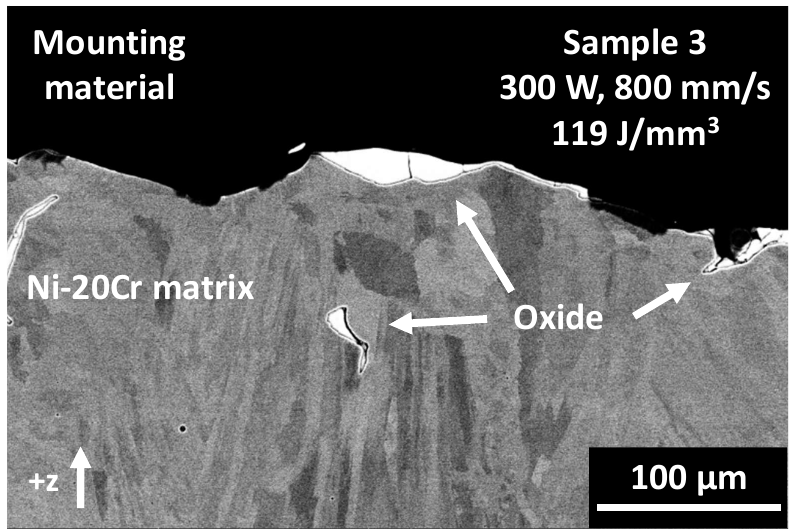}
    \caption{Backscattered electron SEM image from top surface of Sample 3 (300 W, 800 mm/s).}
    \label{fig:topSlag}
\end{figure}

While micron-scale oxide inclusions are observed in all of the experimental samples, the maximum particle diameter predicted by the model is typically less than 500 nm, which suggests that some of the physics of oxide agglomeration are not captured by the current framework. From Fig. \ref{fig:UstructureComparison}(c), it is clear that the area\% of micron-scale oxide inclusions increases strongly with scanning velocity. Kenel \textit{et al.}~observed a similar trend in a Ni-Cr-Al-Ti alloy coated with 0.5 wt.\% $-$ 1 wt.\% Y$_2$O$_3$ nanoparticles, claiming that Al reacted with Y$_2$O$_3$ to form a low-melting point slag that pooled at the top surface of the melt pool \cite{kenel}. Y$_2$O$_3$ is not expected to interact in a similar way with Ni or Cr in the current investigation, yet similar oxide deposits are observed at the top surface of all experimental samples, as shown in Fig. \ref{fig:topSlag} for Sample 3 (300 W, 800 mm/s). \par

\vspace{10px}

These observations indicate that the Y$_2$O$_3$ added to the feedstock melted during processing. This conclusion is supported by the solubility curve of Y$_2$O$_3$ in liquid Ni-20Cr calculated from the modeling framework, which predicts negligible solubility (less than 0.03 wt.\%) of the oxide at its equilibrium melting point (2700 K). Even at the vaporization temperature of Ni (3000 K), the calculated solubility of Y$_2$O$_3$ in the liquid alloy is less than 0.11 wt.\%, which implies that a substantial fraction of the oxide will melt prior to dissolution at higher temperatures. \par

Melting of the oxide may result in accelerated coarsening and agglomeration, which could explain the formation of the bulk inclusions in Fig.~\ref{fig:UstructureComparison}(b). At high scanning velocity, not all of the oxide agglomerates have sufficient time to float to the top of the melt pool, causing them to get trapped in the solid alloy before they have sufficient time to minimize their surface area. The area\% of these micron-scale inclusions decreases strongly at low scanning velocity due to the ejection of oxide-rich spatter from the melt pool (which is more severe at higher laser power and lower scanning velocity) \cite{kenel, haines}.

The high likelihood of oxide melting enables an alternative mechanism for dispersoid formation that is not considered in the current modeling framework. Specifically, since deposits of the liquid oxide can deform under the influence of turbulent melt pool flows, it is possible for these large deposits to break apart into smaller nanometric droplets and get dispersed in the melt pool, as proposed by Zhong \textit{et al.}~\cite{zhong}. Nevertheless, droplets of the liquid oxide (melting point 2700 K) are likely to solidify well in advance of the liquid alloy (melting point 1690 K), so these particles are expected to remain in the solid state for the majority of their residence time in the melt pool, in alignment with the model assumption (Section \ref{assumptions}). Further work is needed to assess the significance of this alternative mechanism for oxide formation, which will likely require a multi-physics process model capturing fluid dynamics.

\subsection{Oxide losses during processing}\label{lossDiscussion}
\begin{figure}[h]
    \centering
    \includegraphics[width=0.5\linewidth]{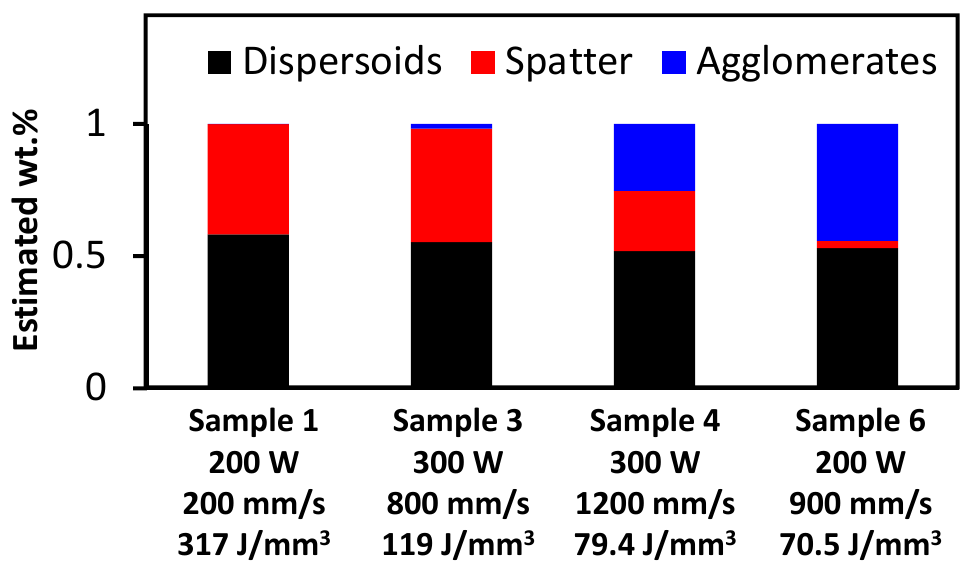}
    \caption{Estimated losses of Y$_2$O$_3$ during processing due to the ejection of oxide-rich spatter from the melt pool and oxide agglomeration.}
    \label{fig:spatterLoss}
\end{figure}

From the preceding section, it is clear that a substantial fraction of the oxide added to the alloy feedstock is lost to agglomeration/spatter during processing, rather than being retained as nanometric dispersoids. Given the estimated vol.\% of oxide agglomerates (micron-scale inclusions) from SEM and the estimated vol.\% of nanometric dispersoids from STEM, oxide losses due to spatter can be approximated, as presented in Fig. \ref{fig:spatterLoss}. \par

Irrespective of the laser power and scanning velocity combination, the estimated concentration of Y$_2$O$_3$ in the printed samples is relatively constant, taking an average value of 0.55 wt.\% across all samples in Fig. \ref{fig:spatterLoss}. While oxide melting and agglomeration during processing are expected to occur for all samples, the energy input determines whether these oxide deposits will be ejected due to spatter (high laser power, low scanning velocity) or incorporated into the bulk microstructure (low laser power, high scanning velocity). The losses of Y$_2$O$_3$ during processing are not as severe as reported by other works, where oxide losses in excess of 70\% for a 316L stainless steel + 1 wt.\% Y$_2$O$_3$ ODS alloy have been observed \cite{zhong}. The Si in the 316L interacted with the added Y$_2$O$_3$ in that study \cite{zhong}, which could have encouraged the formation of a lower-melting point compound that was more prone to spatter and agglomeration than the nominally-pure Y$_2$O$_3$ considered in the current investigation. \par

Since the modeling framework enforces mass conservation, another set of calculations was performed where the concentration of oxide in the system is less than the nominal 1 wt.\% to account for oxide losses during processing. The results of the calculations setting the Y$_2$O$_3$ concentration equal to 0.52 $-$ 0.58 wt.\% (depending on the vol.\% of dispersoids measured with STEM) are presented in Fig.~\ref{fig:solTime} and show improved agreement with the experimental measurements. Since the final number density is limited by collision coarsening, the solidification time has a comparatively stronger influence on the number density than the Y$_2$O$_3$ concentration. On the other hand, the oxide diameters depend on the driving force for diffusional growth in addition to the solidification time, so reducing the concentration of Y$_2$O$_3$ leads to a substantial reduction ($\sim$25\%) in the predicted mean diameter. The results in Fig. \ref{fig:solTime} support the hypothesis that solidification time should be minimized in order to maximize the dispersoid number density. \par

\subsection{Strategies for future experiments}
\begin{figure}[th]
    \centering
    \includegraphics[width=0.9\linewidth]{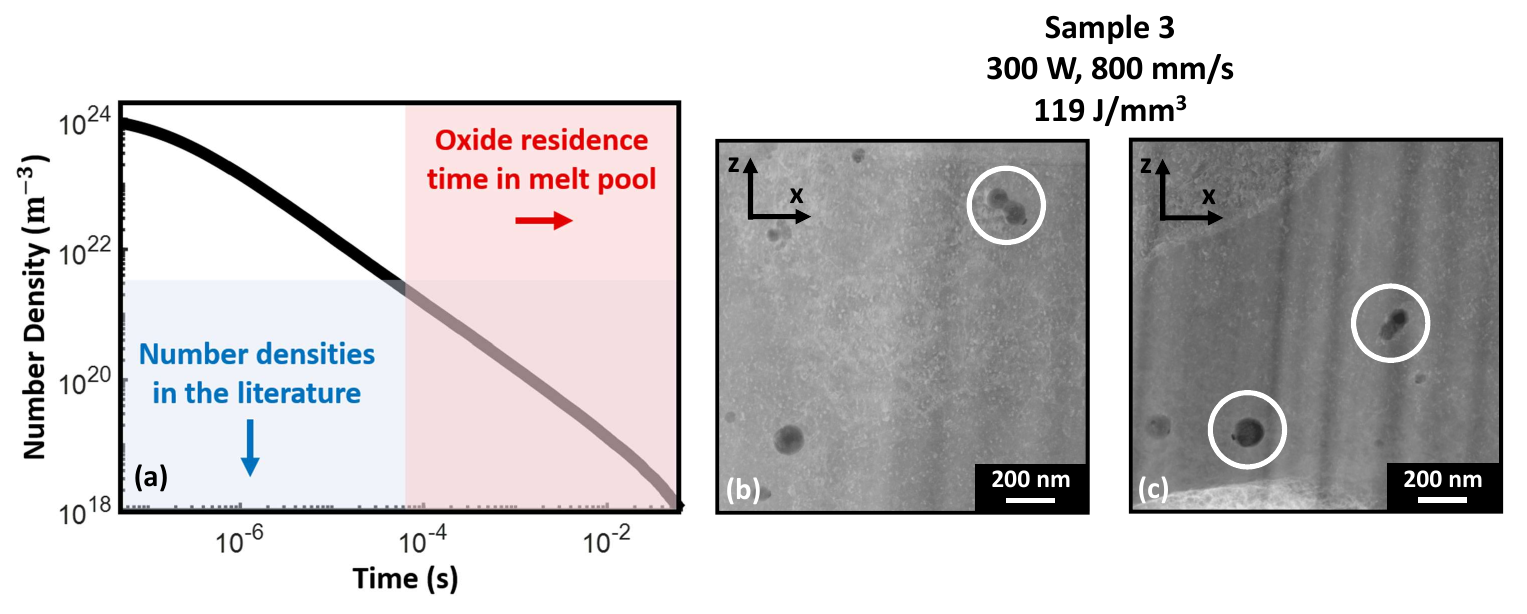}
    \caption{(a) Number density evolution for a monodisperse system. The upper bound for the dispersoid number density (3.4$\times$10$^{21}$ m$^{-3}$) is taken from \cite{horn}. The lower bound for the dispersoid residence time (74 $\mu$s) is taken from \cite{Eo}. (b),(c) STEM images showing potential oxide collisions in the Sample 3 (300 W, 800 mm/s) bulk microstructure.}
    \label{fig:coarsening}
\end{figure}

PBF-LB processing conditions are constrained to a narrow laser power and scanning velocity window where porosity defects can be minimized (Fig. \ref{fig:opticalPVmap}), which places limitations on the accessible range of solidification times. To compare solidification times against coarsening time scales in the melt pool, the particle collision model was used to simulate the evolution of a monodisperse system of critical nuclei, with initial radius 1 nm and initial number density 10$^{24}$ m$^{-3}$ (0.5 vol.\%, which is typical of ODS alloys \cite{odette}). The result (neglecting nucleation and diffusional dissolution/growth) is shown in Fig.~\ref{fig:coarsening}(a). From Fig.~\ref{fig:coarsening}(a), it is clear that solidification in PBF-LB is not sufficiently fast to yield the high number densities achievable with powder metallurgy ($>$10$^{23}$ m$^{-3}$). Figures \ref{fig:coarsening}(b) and \ref{fig:coarsening}(c) present potential empirical evidence of particle collisions, where some dispersoids identified in the STEM images appear to be in the process of merging together. \par

\begin{figure}[t]
    \centering
    \includegraphics[width=0.5\linewidth]{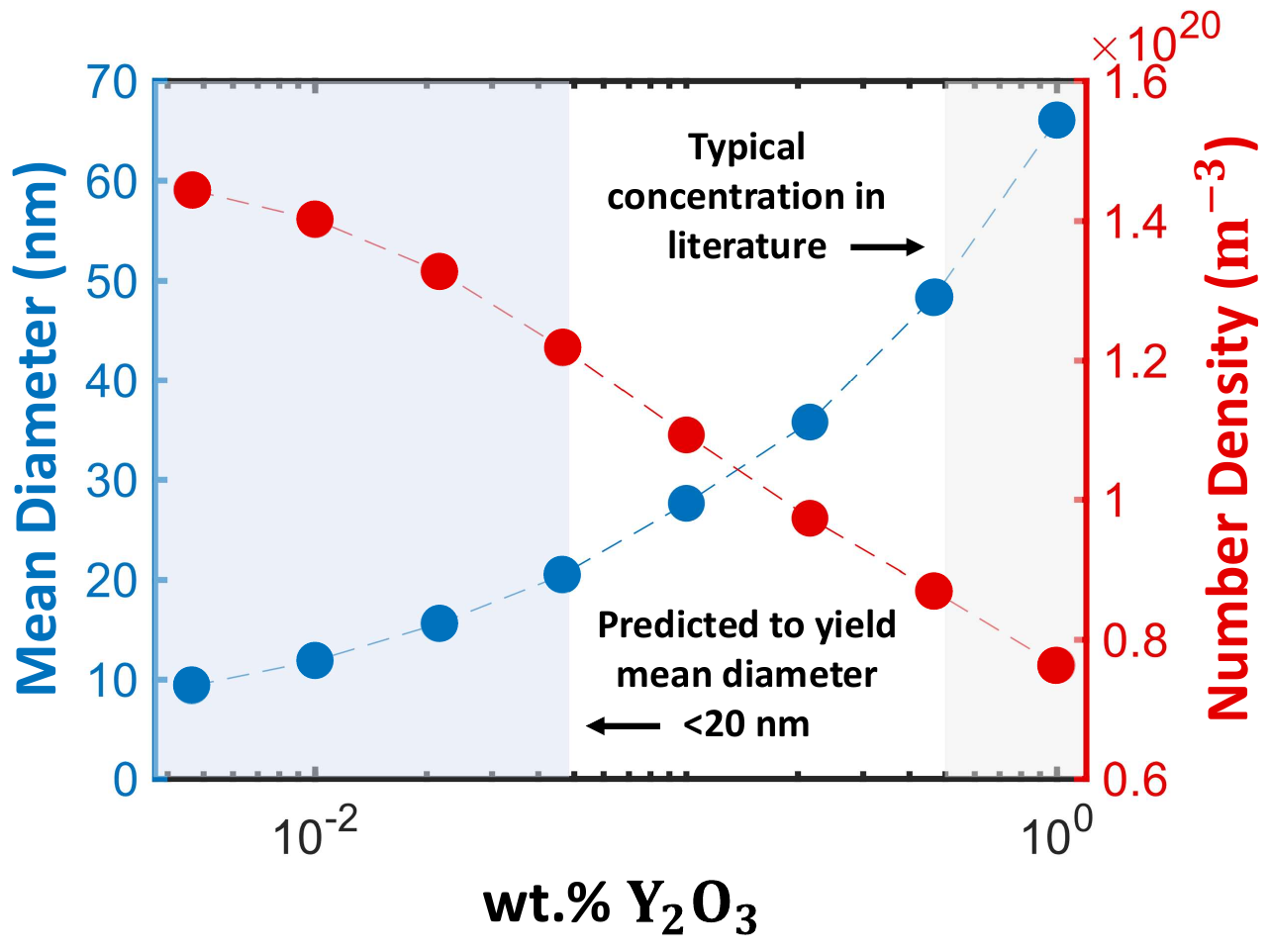}
    \caption{Effect of Y$_2$O$_3$ concentration on final mean diameter and number density. Each vertical pair of data points uses the same temperature history (Sample 4; 300 W, 1200 mm/s), which is shown in Fig. \ref{fig:cameraTemps}(c). Typical Y$_2$O$_3$ concentrations are taken from \cite{nasa20, nasa23, zhong, kenel, boegelein, hunt, spierings}. The optimum mean dispersoid diameter for high-temperature creep resistance is $\sim$20 nm \cite{rosler}.}
    \label{fig:varyWt}
\end{figure}

Since number densities in AM are constrained by collision coarsening, varying the concentration of oxide in the alloy may be an effective way to control dispersoid size, as shown in Fig. \ref{fig:varyWt}. The result in Fig. \ref{fig:varyWt} does not incorporate oxide losses due to agglomeration/spatter, so mean diameters are likely overpredicted for a given wt.\% Y$_2$O$_3$. Nevertheless, it is clear that the optimum mean dispersoid size ($\sim$20 nm for high-temperature creep resistance \cite{rosler}) can be achieved at much lower oxide concentrations than 1 wt.\%. A reduction in oxide concentration may also help mitigate agglomeration during the feedstock coating process and reduce the formation of deleterious micron-scale inclusions. \par

As suggested by the results presented in Figs. \ref{fig:solTime}(b) and \ref{fig:coarsening}, number densities can be increased beyond those achieved in the current investigation by limiting the residence time of oxides in the melt pool. One potential strategy is to reduce the temperature at which oxides nucleate in the melt pool, which is largely a function of the thermodynamic stability of the oxide phase. For example, Eo \textit{et al.}~showed that Si-Mn-Cr-O oxides (with lower thermodynamic stability than Y$_2$O$_3$) form below 1700 K during PBF-LB, which limits their residence time in the melt pool to less than 74 $\mu$s [red shading in Fig. \ref{fig:coarsening}(a)]. High thermodynamic stability provides favorable resistance to Ostwald ripening under service conditions, so this strategy may not be suitable for all applications. A similar effect could be achieved by using an alloy matrix with a high melting point (e.g., refractories), though this could come at the cost of processability. \par

Another strategy to limit oxide residence time in the melt pool is to leverage the absorption of atmospheric oxygen into the melt pool to drive oxide formation, as investigated in prior works \cite{horn, haines}. This approach could help to increase the oxygen concentration in low temperature regions of the melt pool, near the advancing solidification front. If the oxygen concentration is sufficiently high to drive oxide nucleation in those regions, then the particles could be quickly incorporated into the nearby solid before they have sufficient time to coarsen significantly. Using this approach with a low energy input (90 W, 1500 mm/s), Horn \textit{et al.}~achieved number densities sufficiently high for high-temperature creep applications [3.4$\times$10$^{21}$ m$^{-3}$, blue shading in Fig. \ref{fig:coarsening}(a)] \cite{horn}. However, despite the relatively high vol.\% of oxides in their samples (estimated to fall between 0.5 vol.\% and 1.2 vol.\%), the measured number densities were still considerably lower than desired for nuclear fusion applications ($>$10$^{23}$ m$^{-3}$). \par

It is important to note that, while the current investigation emphasizes the as-fabricated dispersoid size and number density, these are not the only characteristics that determine the suitability of ODS alloys for final applications. For example, for nuclear reactor applications, further work is needed to understand the long-term stability of the dispersoids under neutron/ion irradiation, which are likely to depend on the dispersoid morphology, dislocation structure, alloy composition, and irradiation conditions. \par

\section{Conclusions}
\begin{itemize}
  \item PBF-LB experiments were combined with a thermodynamic/kinetic modeling framework to understand the evolution of Y$_2$O$_3$ nanoparticles during processing. The evolution can be divided into three phases: (I) dissolution upon heating; (II) nucleation and diffusional growth upon cooling; and (III) growth via collision coarsening (Fig. \ref{fig:oxideEvolution}).
  \item In the fabricated samples, the smallest mean dispersoid diameter (29 nm) and highest number density (1.0$\times$10$^{20}$ m$^{-3}$) were observed at 300 W and 1200 mm/s. The largest mean diameter (72 nm) and lowest number density (1.5$\times$10$^{19}$ m$^{-3}$) were observed at 200 W and 200 mm/s.
  \item Smaller dispersoid sizes and higher number densities were correlated with shorter solidification time scales. The results of the modeling framework agreed with this trend.
  \item A considerable fraction of the oxide added to the feedstock was lost during processing, due to oxide agglomeration and the ejection of oxide-rich spatter from the melt pool.
  \item The model suggests that the mechanism that limits the final number density is collision coarsening of dispersoids in the melt pool [Fig.~\ref{fig:coarsening}(a)].
\end{itemize}

\section{Acknowledgements}
Dr. Sudarsanam Babu (University of Tennessee, Knoxville) is acknowledged for technical discussions. Scott Kram (CMU Mill 19) is acknowledged for his help with the PBF-LB experiment. SEM was performed at the Materials Characterization Facility at CMU. The metal powder used in this publication was provided to the authors under Space Act Agreement SAA1-32576/PAM 32576 with the National Aeronautics and Space Administration. The support is gratefully acknowledged. L.S. acknowledges the partial support from the National Nuclear Security Administration (NNSA) grant DE-NA0003921.

\newpage
\appendix

\section{Build setup}
\label{sec:buildSetup}

\begin{figure}[h]
    \centering
    \includegraphics[width=0.8\linewidth]{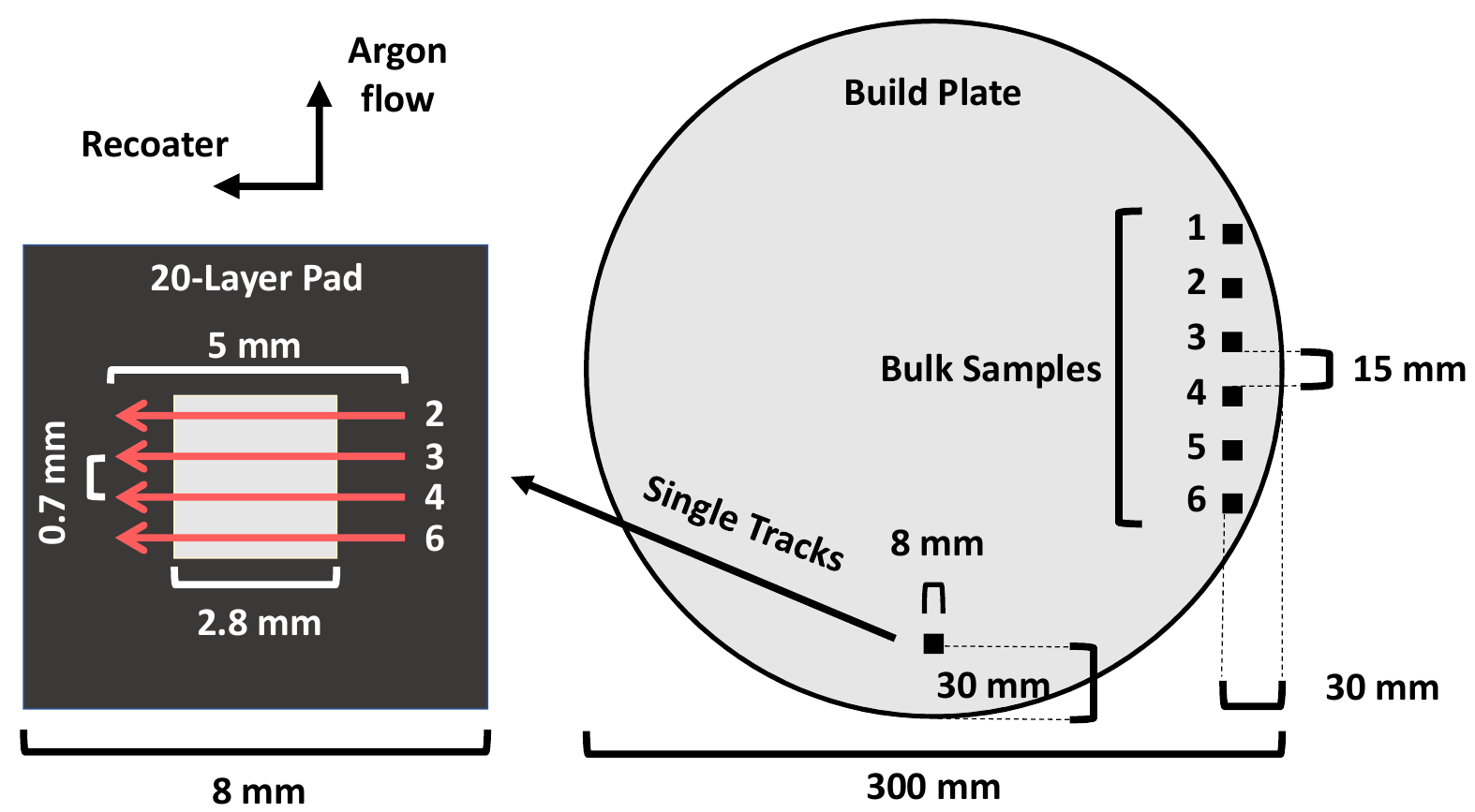}
    \caption{A schematic of the build setup in the PBF-LB machine. The red arrows show the locations of the single tracks. The light-colored square inset on the 20-layer pad shows the field of view of the camera (2.8 $\times$ 2.8 mm$^2$). The numbers next to the bulk samples and single tracks correspond to the sample numbers in Table \ref{tab:PVcombos}.}
    \label{fig:buildSchematic}
\end{figure}

The quantity of feedstock powder available for the experiment was limited. Therefore, the bulk samples were placed on one side of the build plate (Fig. \ref{fig:buildSchematic}), close to the recoater entrance. This arrangement mitigates the risk of defects arising from powder short-spreading. The thin pad and single tracks were positioned in a specific location near the front of the build plate within the camera's field of view. To allow for sufficient spacing between the single tracks, only four tracks could be deposited within the camera's field of view in a single experiment. \par

\section{Construction of composite thermal image}
\label{sec:compositeConstruction}

\begin{figure}[th]
    \centering
    \includegraphics[width=0.99\linewidth]{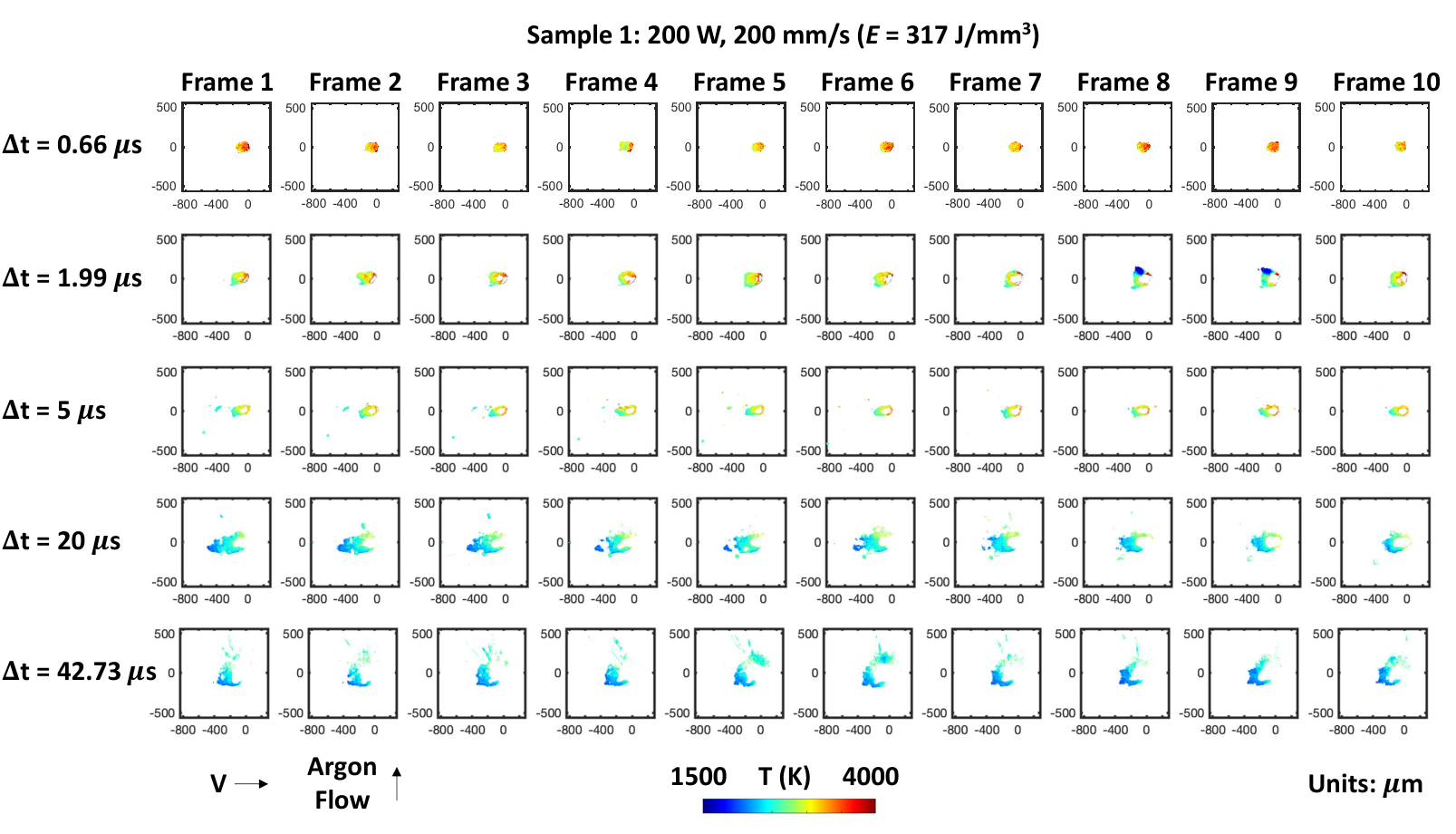}
    \caption{Individual thermal images of the top surface of the melt pool for Sample 1 (200 W, 200 mm/s) for different exposure times.}
    \label{fig:compositeConstruction}
\end{figure}

The construction of the composite temperature field at the top surface of the melt pool is demonstrated in Fig. \ref{fig:compositeConstruction}. In the figure, it is clear that the temperature field in the melt pool is not constant over time (i.e., the thermal image changes between frames for a given exposure time). Averaging the measurements across ten different frames helps to reduce the influence of temporal variations in the melt pool temperature field. Additionally, fitting a polynomial to the raw temperature measurements helps to smooth over noise in the thermal images (Fig. \ref{fig:cameraTemps}). \par

Since each track was re-scanned multiple times to record videos at varying exposure times, heat accumulation could influence the temperature field within the melt pool. Though the deposition temperature of the single tracks may have varied between exposure times, the influence of these variations is reduced by averaging the pixel values in the composite images. Since the temperature histories for the modeling framework were extracted from the composite images, the effects of heat accumulation in the single tracks are not expected to contribute meaningfully to the uncertainty in the model predictions. \par

\section{Oxide composition}
\label{sec:ellingham}

\begin{figure}[h]
    \centering
    \includegraphics[width=0.5\linewidth]{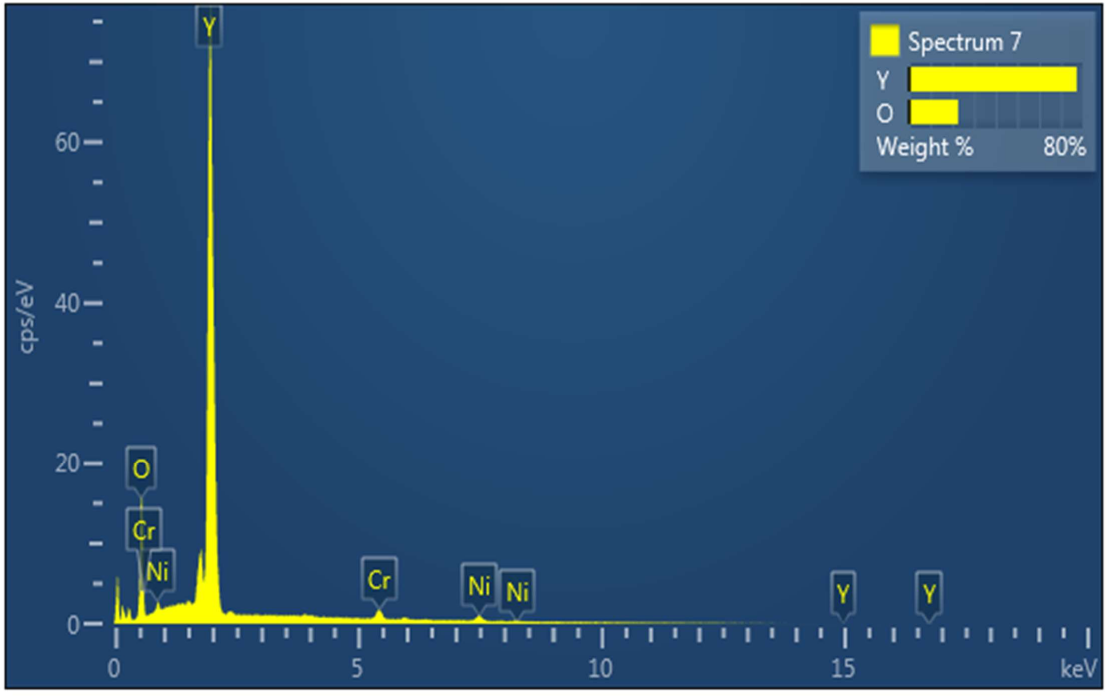}
    \caption{Representative EDS spectrum extracted from an oxide inclusion in Sample 3 (300 W, 800 mm/s).}
    \label{fig:edsSpectrum}
\end{figure}

\begin{figure}[h]
    \centering
    \includegraphics[width=0.5\linewidth]{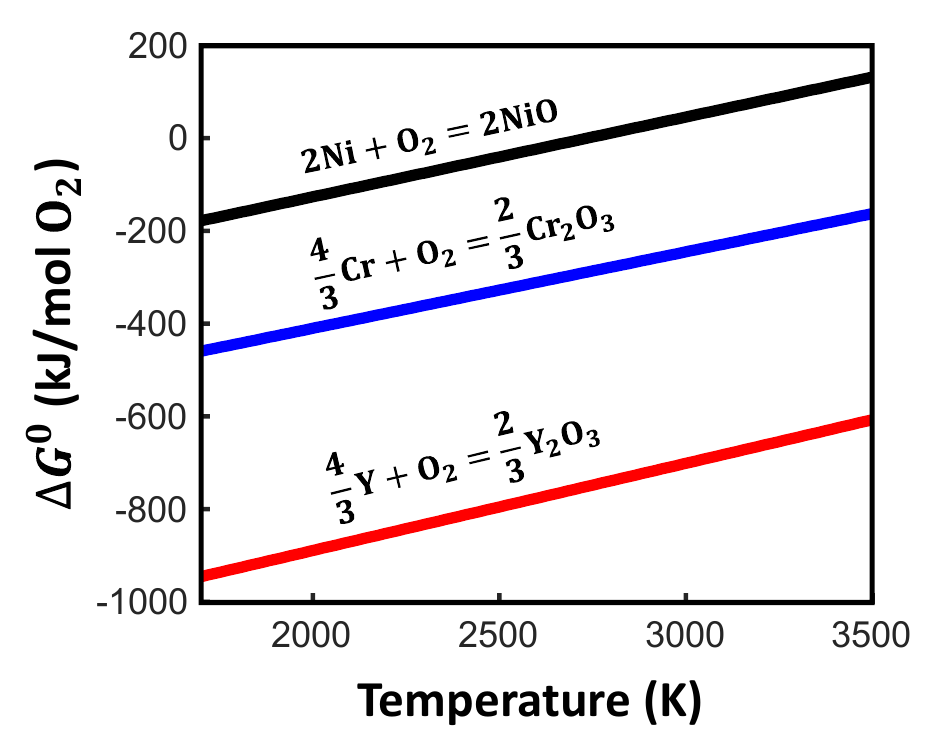}
    \caption{Ellingham diagram for Ni-20Cr + 1 wt.\% Y$_2$O$_3$ alloy. Gibbs free energy expressions from \cite{turkdogan1980physical}, normalized to 1 mol O$_2$.}
    \label{fig:ellingham}
\end{figure}

EDS spectra of the oxide inclusions were obtained from a variety of different locations in all of the fabricated samples. A representative spectrum from an inclusion in Sample 3 is provided in Fig. \ref{fig:edsSpectrum}. From the Ellingham diagram in Fig. \ref{fig:ellingham}, it is clear that Y is the strongest deoxidizer in the Ni-20Cr ODS alloy. Therefore, any dissolved O in the melt pool will react preferentially with Y before Cr or Ni, assuming the solute concentrations of both Y and O are sufficiently high to drive the deoxidiation reaction. The build chamber was maintained at less than 0.1\% O$_2$ during the PBF-LB experiment. Therefore, the only significant source of oxygen in the melt pool is the added Y$_2$O$_3$. As a consequence, it is likely that the composition of dissolved Y and O in the melt pool will follow the stoichiometry of Y$_2$O$_3$, preventing the formation of less-stable oxides.

\newpage
\section{Analysis of model assumptions}
\label{sec:analyzeAssumptions}
\begin{figure}[hbt]
    \centering
    \includegraphics[width=0.8\linewidth]{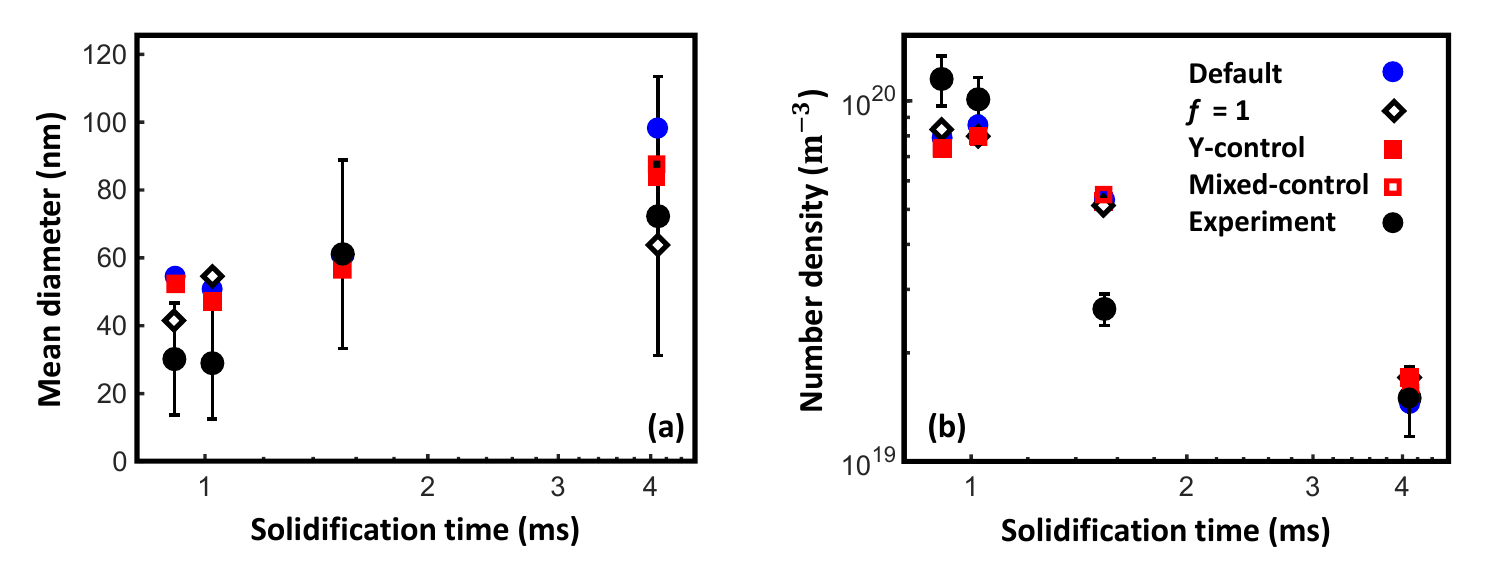}
    \caption{Comparison of model results under a variety of different assumptions. The experimental data and default model result are reproduced from Fig. \ref{fig:solTime} in the main text. Error bars are not included on the model results for clarity. (a) Comparison of dispersoid mean diameters. The experimental vertical error bars represent one standard deviation across all measured diameters for a given sample. (b) Comparison of dispersoid number densities. The experimental vertical error bars are the standard error in number density across all micrographs for each sample.}
    \label{fig:assumptionComparison}
\end{figure}

Figure \ref{fig:assumptionComparison} presents model results using 0.52 $-$ 0.58 wt.\% Y$_2$O$_3$. The default case assumes O-controlled growth and non-unity activity coefficients [calculated from Eqs. \eqref{eq:activityY} and \eqref{eq:activityO}], as presented in Fig. \ref{fig:solTime} of the main text. The model trends are unaffected by the choice of growth mechanism (O-controlled, Y-controlled, or mixed). Setting the activity coefficients ($f$) equal to unity does not strongly affect the final number density, but the mean diameter becomes less sensitive to the solidification time. For a given temperature, the concentration of oxygen at the oxide-alloy interface ($c^i_O$) increases with decreasing $f$ [Eq. \eqref{eq:eqbm}]. Consequently, setting $f$ to unity slows the rate of diffusional dissolution and growth [since the magnitude of $c^*$ is smaller, Eq. \eqref{eq:cstar}]. This slowed growth manifests as a smaller final mean diameter at the longest solidification time (4.1 ms).

\begin{figure}[th]
    \centering
    \includegraphics[width=1\linewidth]{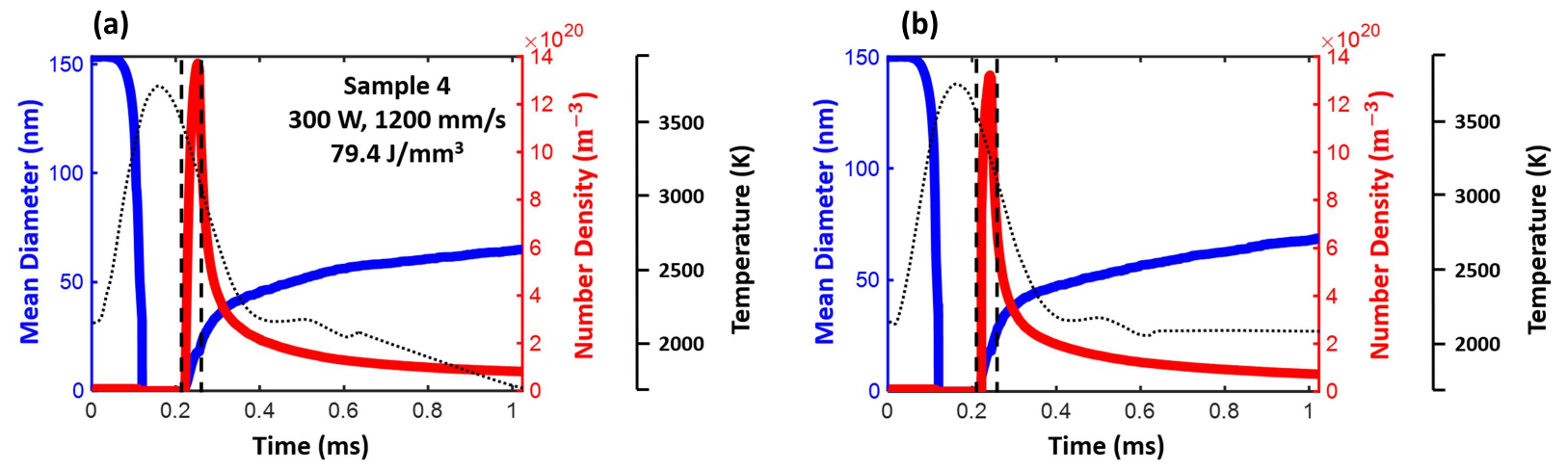}
    \caption{(a) Predicted evolution of the mean diameter and number density of 1 wt.\% Y$_2$O$_3$ nanoparticles for the Sample 4 (300 W, 1200 mm/s) temperature history with linear cooling to the alloy melting point (1690 K). (b) Predicted evolution with a constant temperature of 2069 K, the final temperature measurement provided by the camera.}
    \label{fig:finalCoolingSensitivity}
\end{figure}

As discussed in Section \ref{sec:surfTempResults}, the temperature histories used as model inputs assume linear cooling to the alloy melting point (1690 K) to compensate for the limited dynamic range of the camera. However, due to the release of latent heat, it is also possible that the tail of the melt pool remains at an elevated temperature compared to the assumed linear cooling profile. These two scenarios are compared in Fig. \ref{fig:finalCoolingSensitivity}. \par

From Fig.~\ref{fig:oxideEvolution}, it is clear that the linear cooling period falls well within the phase of oxide evolution dominated by collision coarsening. The time scales for these collisions are shorter at higher temperatures, so the final number density is 8\% lower in Fig. \ref{fig:finalCoolingSensitivity}(b) compared to Fig. \ref{fig:finalCoolingSensitivity}(a) and the final mean diameter is 2\% higher. Nevertheless, it is clear that uncertainty in the melt pool temperature does not significantly affect the oxide evolution in the collision coarsening regime.

\section{Nucleation sensitivity analysis}
\label{sec:nucleationSensitivity}
\begin{figure}[h]
    \centering
    \includegraphics[width=0.45\linewidth]{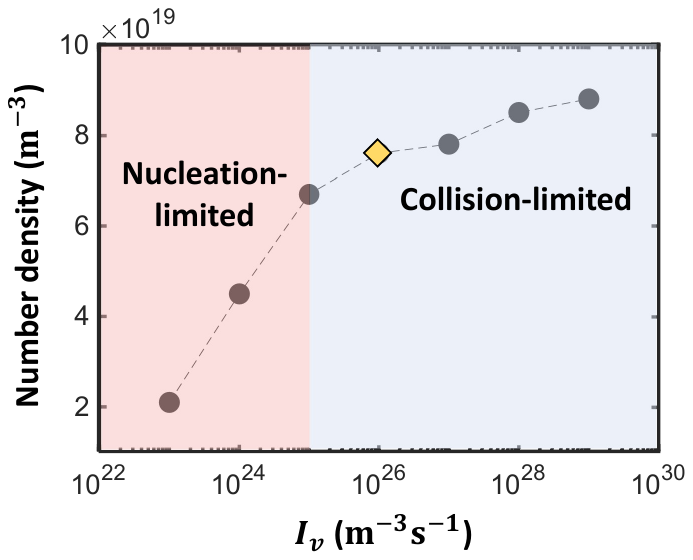}
    \caption{Influence of the assumed nucleation rate ($I_v$) on the final dispersoid number density. For the results in the main text, $I_v$ was held constant at 10$^{26}$ m$^{-3}$s$^{-1}$, which is indicated by the yellow diamond.}
    \label{fig:nucleationRate}
\end{figure}

Calculations were performed to evaluate the sensitivity of the model to the assumed values of $I_v$, as shown in Fig. \ref{fig:nucleationRate}. At low values of $I_v$ ($<$10$^{25}$ m$^{-3}$s$^{-1}$), the final number density is limited by the total number of nuclei that form upon cooling. During this nucleation-limited regime, the final number density increases strongly with $I_v$. At higher values of $I_v$ ($>$10$^{25}$ m$^{-3}$s$^{-1}$), the number density increases rapidly over a short nucleation window (Fig. \ref{fig:oxideEvolution}), but the final value is ultimately limited by collision coarsening. Consequently, in this collision-limited regime, the final number density increases weakly with $I_v$. The overall trends in dispersoid size and number density predicted from the modeling framework are unaffected by the choice of $I_v$ within the collision-limited regime. \par

\newpage

\begin{figure}[h]
    \centering
    \includegraphics[width=0.7\linewidth]{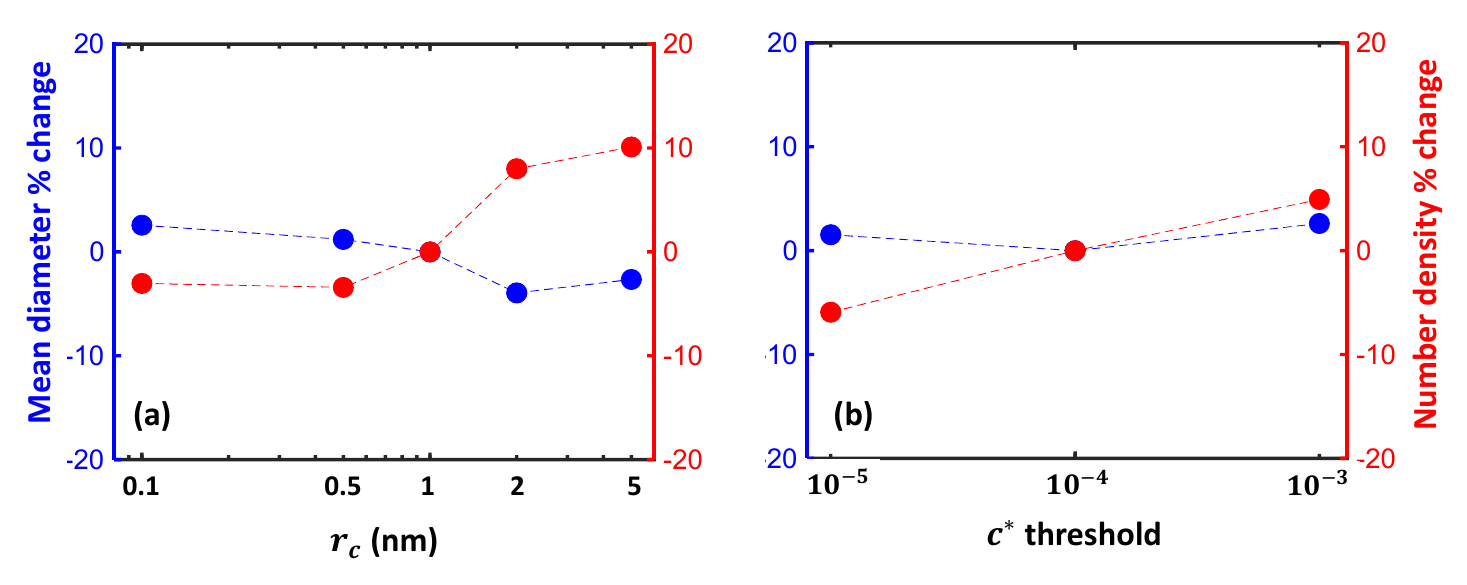}
    \caption{All results use the same temperature history, Sample 4 (300 W, 1200 mm/s), shown in Fig. \ref{fig:cameraTemps}(c) of the main text. (a) Influence of the assumed value of the Y$_2$O$_3$ critical radius ($r_c$) on the final mean diameter and number density of dispersoids. (b) Influence of the $c^*$ nucleation threshold on the final mean diameter and number density.}
    \label{fig:rc_cstarSensitivity}
\end{figure}

Figures \ref{fig:rc_cstarSensitivity}(a) and \ref{fig:rc_cstarSensitivity}(b) show the sensitivity of the model to the assumed Y$_2$O$_3$ critical radius and $c^*$ threshold, which controls whether new nuclei are added to the system in a time step. For the results in the main text, $r_c$ is set constant at 1 nm and the $c^*$ threshold is set to 1$\times$10$^{-4}$. Neither parameter has a significant impact on the final mean diameter or number density ($<$10\% change compared to the results in the main text). The lack of sensitivity to the $c^*$ threshold results from the high cooling rates in PBF-LB. During cooling, the value of $c^*$ increases with decreasing temperature for a given bulk O concentration. Therefore, if the temperature rapidly drops, $c^*$ may increase by several orders of magnitude within a short time span (typically $<$10\% of the overall solidification time). As a result, the timing of the nucleation event upon cooling is not significantly affected by the $c^*$ threshold, and the oxide residence times are relatively constant for all threshold values.

\section{Solidification time estimation}
\label{sec:poolLength}
Heigel and Lane performed single track laser scans on Alloy 625 substrates to measure the melt pool length for a variety of laser power and scanning velocity combinations \cite{heigel2018measurement}. Two of those combinations are similar to Sample 1 (200 W, 200 mm/s) and Sample 6 (200 W, 900 mm/s) in the current investigation. These experimental results are reproduced in Table \ref{tab:heigelResults}. 

\begin{table}[h]
    \centering
    \small
    \caption{Experimental melt pool length measurements from \cite{heigel2018measurement}.}
    \label{tab:heigelResults}
    \begin{tabular}{ccc}
        \toprule
        \textbf{Laser power (W)}&\textbf{Scanning velocity (mm/s)}&\textbf{Melt pool length ($\boldsymbol{\mu}$m)}\\
        \midrule
        195&200&824$\pm$109\\
        195&800&813$\pm$79\\
        \bottomrule
    \end{tabular}
\end{table}

Heigel and Lane concluded that the melt pool length is independent of the scanning velocity when the laser power is sufficiently high ($>$195 W), in alignment with the analytical expression derived from the Rosenthal equation \cite{rosenthal1941mathematical}. 
\vspace{10px}
\begin{equation}
\label{eq:length}
    {L = \frac{\alpha P}{2\pi k (T_{melt} - T_{\infty})}},
\end{equation}

\noindent where $L$ is melt pool length, $\alpha$ is the laser absorptivity, $P$ is laser power, $k$ is thermal conductivity, $T_{melt}$ is the alloy melting temperature, and $T_{\infty}$ is the far-field temperature. Therefore, the two measurements in Table \ref{tab:heigelResults} are averaged to obtain a representative value of the melt pool length for Sample 1 (200 W, 200 mm/s) and Sample 6 (200 W, 900 mm/s), equal to 819 $\mu$m. The uncertainties are also averaged, equal to 94 $\mu$m, which is 11\% of the average length. Following Eq.~\eqref{eq:length}, the average length at 200 W (819 $\mu$m) is multiplied by 1.5 to obtain a representative length for Sample 3 (300 W, 800 mm/s) and Sample 4 (300 W, 1200 mm/s), equal to 1230 $\mu$m. An uncertainty of 11\% is assumed, taking a value of 141 $\mu$m. All of these lengths are divided by the corresponding scanning velocity to obtain the estimated solidification times and uncertainties reported in Fig. \ref{fig:cameraTemps} of the main text. 

 \bibliographystyle{elsarticle-num} 
 \bibliography{refs}





\end{document}